\documentclass[a4paper,11pt]{article}
\pdfoutput=1
\usepackage{amssymb}
\usepackage{amsmath}
\usepackage{mathtools}
\usepackage[usenames,dvipsnames,svgnames,table]{xcolor}
\usepackage[utf8]{inputenc}
\usepackage{color}
\usepackage{cite}
\usepackage{subcaption}
\usepackage{lmodern}
\usepackage{footnote}
\usepackage[normalem]{ulem}
\usepackage{jheppub}
\usepackage{hyperref}
\usepackage{verbatim}
\usepackage{listings}
\usepackage{dsfont}
\usepackage{nicefrac,xfrac}
\newcommand{\ThreeParticleCutoff}[0]{Hansen:2014eka}
\newcommand{\leadingEnergyExpansion}[0]{Huang:1957im,Luscher:1986pf,Beane:2007qr,Hansen:2016fzj,Grabowska:2021xkp}
\newcommand{\DiscEffectsOtherContext}[0]{Draper:2023boj}
\newcommand{\ThreeBodyPapers}[0]{Polejaeva:2012ut,Hansen:2014eka,Hansen:2015zga,Hammer:2017uqm,Hammer:2017kms,Mai:2017bge,Doring:2018xxx,Hansen:2019nir}
\newcommand{\LeftHandCut}[0]{Raposo:2023oru,Du:2023hlu,Padmanath:2022cvl,Meng:2021uhz,Green:2021qol}
\newcommand{\heavyquark}[0]{Mohler:2012na,Mohler:2013mhk,Lang:2014yfa,Gayer:2021xzv,Lang:2022elg,Bulava:2023wrz,Yeo:2024chk}

\renewcommand{\vec}[0]{\boldsymbol}

\renewcommand{\>}{\rangle}
\newcommand{\f}{{{\sf f}}}
\newcommand{\Z}{{{\sf Z}_a}}
\newcommand{\<}{\langle}

\newcommand{\Fs}{\boldsymbol{F}^{\sf S}_{\!a}}
\newcommand{\deltaFs}{\delta \boldsymbol{F}^{\sf S}_{\!a}}
\newcommand{\deltaFsPV}{\delta \boldsymbol{F}^{\sf S, pv}_{\!a}}
\newcommand{\Ms}{\mathcal M^{\sf S}_{a}}

\definecolor{jlab_red}{RGB}{192,39,45}
\definecolor{jlab_orange}{RGB}{249,102,0}
\definecolor{jlab_blue}{RGB}{47,122,121}
\definecolor{jlab_green}{RGB}{65,125,10}
\definecolor{jlab_blue}{RGB}{47,122,121}

\hypersetup{%
pdftitle = {discretization-scattering},
pdfsubject = {QCD},
pdfauthor = {Hansen and Peterken},
colorlinks = {true},
filecolor = {black},
linkcolor = {jlab_blue},
menucolor = {black},
citecolor = {jlab_blue},
urlcolor = {jlab_blue},
}{}

\title{Discretization effects in finite-volume $2\to2$ scattering}

\author[a]{M.~T.~Hansen}
\author[a]{and T.~Peterken}

\affiliation[a]{Higgs Centre for Theoretical Physics, School of Physics and Astronomy, University of Edinburgh, Edinburgh EH9 3FD, UK}
\emailAdd{maxwell.hansen@ed.ac.uk}
\emailAdd{t.peterken@sms.ed.ac.uk}

\abstract{We incorporate non-zero lattice-spacing effects into L{\"u}scher's finite-volume scattering formalism. The new quantization condition takes lattice energies as input and returns a version of the discretized scattering amplitude whose definition is transparent in the context of Symanzik Effective Theory. In contrast to the standard formalism, this approach uses single-hadron discretization effects to define modified versions of the finite-volume zeta functions. The new formalism requires two sets of angular-momentum indices, which encode the ultraviolet mixing of angular momentum states (due to the lattice spacing), in addition to the well-known infrared mixing (due to the finite volume).}

\begin{document}

\maketitle
\flushbottom
\abovedisplayskip 11pt
\belowdisplayskip 11pt

\section{Introduction}

The study of scattering amplitudes is a central problem in quantum field theory. While great progress has been made in perturbative calculations of scattering amplitudes in various theories, there are many systems for which such methods do not apply. An important example is the scattering of low-energy degrees of freedom, called hadrons, in quantum chromodynamics (QCD), the quantum theory of the strong force. QCD is most simply expressed in terms of its high-energy degrees of freedom, quarks and gluons, and no analytical methods are known for relating the defining Lagrangian directly to hadronic amplitudes.

Over the last decades, lattice QCD has increasingly established itself as a reliable method to address this challenge numerically. The approach is based on defining the quantum path integral on a discretized, finite-volume Euclidean spacetime and using Monte Carlo importance sampling to numerically estimate Euclidean correlators. In the context of scattering, such correlators can be used to numerically extract finite-volume energy levels. Then, using the L{\"u}scher formalism~\cite{Luscher:1986pf} and extensions~\cite{Rummukainen:1995vs,He:2005ey,Christ:2005gi,Kim:2005gf,Lage:2009zv,Bernard:2010fp,Fu:2011xz,Briceno:2012yi,Hansen:2012tf,Guo:2012hv,Briceno:2014oea}, these finite-volume energies are related to infinite-volume scattering amplitudes via mathematical identities called quantization conditions.\footnote{See ref.~\cite{Briceno:2017max} for a recent review. An alternative, not discussed further in this paper, is the HAL-QCD method~\cite{Murakami:2022cez,Aoki:2023nzp}. This approach defines a quantum-mechanical potential that can be numerically estimated and subsequently used to define a Schr\"odinger equation, the solution of which encodes the hadronic scattering phase shift.}

Quantization conditions are typically derived in the framework of a finite-volume, \emph{continuum} quantum field theory.\footnote{See, however, ref.~\cite{\DiscEffectsOtherContext} for discretization effects applied in the context of finite-volume scattering relations.}
As a result, the relations strictly apply for data in which one takes the continuum (lattice-spacing-to-zero) limit at fixed physical volume. This requires matching physical volumes at different lattice spacings $a$, which is challenging in practice. As a result, all calculations to date have instead applied the quantization condition directly to finite-volume data at fixed $a$. This procedure is justified, for example, when the statistical uncertainties arising from the Monte Carlo sampling are larger than the discretization effects.

An alternative perspective is to note that using the continuum formalism with lattice-shifted finite-volume energies simply gives a version of the scattering amplitude that also contains lattice artifacts. One might then aim to parameterize and extrapolate away such deviations at the level of scattering parameters, without the need of matching physical volumes at different lattice spacings. This is challenging, due to the lacking theoretical understanding of the $a$-dependence, which is governed by the details of the lattice action, as well as the scattering channel(s) under consideration, the kinematics of the scattering particles (including the total spatial momentum in the finite-volume frame), and the choice of how lattice artifacts are treated in the single-hadron masses entering the formalism.

\begin{figure}
\centering
\includegraphics[width=0.9\textwidth]{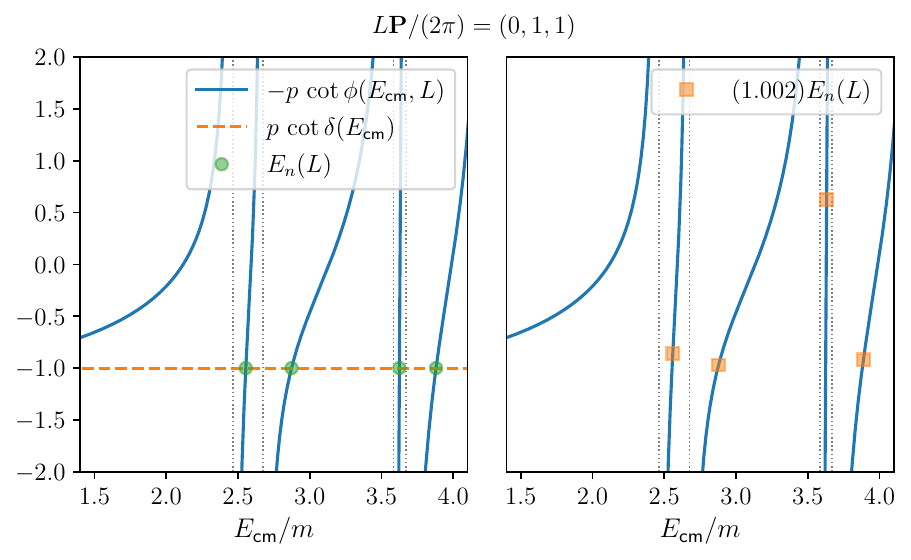}
\caption{Toy example illustrating the potential importance for carefully treating discretization effects in lattice QCD scattering calculations. In both panels the solid blue curve shows the finite-volume function introduced in ref.~\cite{Rummukainen:1995vs} for relating the scattering observable, $p \cot \delta$, to energies extracted with momentum $\boldsymbol P^2 = 2(2\pi/L)^2$ in the finite-volume frame, with $mL=5$, where $m$ is the mass of the scatterer. {\em Left panel}: The orange dashed curve represents a simple scenario for the scattering phase shift and the intersections, marked as green circles, represent the energies that would be extracted in an infinite-precision continuum calculation. {\em Right panel}: The orange squares represent the incorrect extraction if the energies are biased to have values that are larger by two per mille. Note that the issue here is not the non-linearity of the finite-volume function but the fact that some branches are quite steep. We emphasize that whenever statistical uncertainties are larger than discretization effects on the energies, then these will be enhanced such that the result is unbiased.}
\label{fig:warm_up}
\end{figure}

A related point is that quantization conditions generally depend on rapidly varying zeta functions, which can potentially enhance discretization effects. As we illustrate in the toy example of figure~\ref{fig:warm_up}, when the continuum finite-volume energy is near a pole or a zero of the zeta-function, a small shift due to discretization effects could significantly modify the extracted scattering parameters. Another delicate example is the case of a shallow bound state, with mass $M_B$ close to a multi-hadron threshold with energy $E_{\sf thresh}$. In this case, the binding energy $E_B = E_{\sf thresh}-M_B$ can be sensitive to discretization effects due to large cancellations. Recent work by the Mainz group has shown such effects in the determination of the binding energy of the H dibaryon~\cite{Green:2021qol}. Observations like this, together with the increasingly stringent precision goals of the field, imply that it is very timely to consider other strategies for including non-zero $a$-effects in the finite-volume analysis.

In this work, we derive an alternative quantization condition, which incorporates lattice spacing effects into the finite-volume functions entering the quantization conditions. Our key proposal is to use single-hadron Euclidean correlators to describe momentum-dependent $a^2$ effects in terms of low-energy parameters, and then to modify the quantization condition with these additional inputs. The resulting scattering amplitude then sits on a conceptually clear trajectory to the continuum limit which may be easier to describe using polynomials and other simple parametrizations, and which can additionally be investigated using effective field theory and related methods. Perhaps most importantly, the framework ensures that the Lorentz-invariant part of the scattering amplitude, extracted using different total momenta in the finite-volume frame, receives the same $a^2$-effects and can thus be consistently fit to various models or parametrizations at fixed $a$.

Restricting attention to a single two-particle channel of identical particles with no intrinsic spin, our approach to derive the discretized quantization condition closely follows the continuum derivations of refs.~\cite{Luscher:1986pf,Kim:2005gf,Hansen:2012tf,Briceno:2014oea}. The starting point is an all-orders expansion of a generic relativistic quantum field theory defined in a finite volume. Instead of a generic continuum theory, we use an improved Symanzik Effective Theory~\cite{Symanzik:1983dc,Bar:2004xp}, which incorporates discretization effects using irrelevant operators with coefficients scaling as $a^2$ and higher powers. Because the lattice breaks the continuous $O(3)$ rotational symmetry, the $a^2$ terms of the Symanzik Effective Theory need not be $O(3)$ invariant, leading to Feynman rules that are clearly distinct from the usual continuum derivation.

To this point we note that, even for the continuum theory studied in refs.~\cite{Luscher:1986pf,Kim:2005gf,Hansen:2012tf,Briceno:2014oea} the cubic finite volume breaks rotational symmetry, manifesting as the well-known mixing of angular momentum states in the standard quantization conditions. New to this work is a second type of mixing, which is $a^2$-suppressed. Both the infrared (volume-induced) and ultraviolet ($a$-induced) coupling of different angular momenta are encoded in our modified finite-volume functions, given in eq.~\eqref{eq:f_disc_final}. Conveniently, the symmetry group is not modified by the discretization, so the standard technology of projecting to irreducible representations of the cubic group can be applied as in the continuum finite-volume case.

The main result of this work, given in eq.~\eqref{eq:final_QC}, takes the familiar form of a quantization condition but with the alternative geometric function of eq.~\eqref{eq:f_disc_final} and with an $a$-dependent version of the scattering amplitude. The matrix space of the new formalism carries two sets of angular-momentum indices which encode the infrared and ultraviolet mixing effects. It follows that the scattering amplitude is also promoted to carry two sets of angular momentum indices, whereby all non-trivial components in the new angular-momentum space are $a^2$ suppressed. As a result, the new index set drops out in the $a \to 0$ limit, so that the standard quantization condition is recovered.

We emphasize that the modifications entering this work result entirely from lattice artifacts in single-hadron states. Therefore, we provide no new prescription in the case that the latter cannot be statistically resolved. These effects break into two categories (i) our framework requires the use of the single-hadron zero-momentum ground state extracted at the same lattice spacing as the finite-volume scattering energies and (ii) additional lattice effects are incorporated by fitting the momentum dependence of the single-hadron states. A simple scenario might be that only the former $a$-dependence can be resolved, in which case this work gives a motivation for the standard approach of extracting the amplitude using discretized rather than continuum masses. We stress here that it is generically impossible to infer $a \to 0$ information from a fixed $a$ lattice ensemble. The advantage of our method is that the resulting scattering amplitude has a theoretically clean definition at finite $a$.

This work applies to discretizations such as improved Wilson and domain-wall fermions, for which the Symanzik Effective Theory does not require additional flavors. In particular, the formalism does not apply to staggered quarks. An extension to accommodate this set-up would clearly be of great interest and is left as future work.

The rest of this paper is organized as follows: Section~\ref{sec:review_cont_deriv} gives a recap of the derivation for the standard (continuum) quantization condition, emphasizing the steps that need to be modified to include discretization effects. Section~\ref{sec:disc_deriv} gives a brief review of Symanzik Effective Theory as a framework for including lattice artifacts in a continuum formulation, and then uses this to modify the derivation of the quantization condition to include such effects. The main result is also presented in section~\ref{sec:disc_deriv}. Then, in section~\ref{sec:simplified_results}, we study the weakly interacting limit and also show how the result may be used in practice, using $\lambda \phi^4$ to illustrate the general approach. We briefly conclude in section~\ref{sec:conclusion} and in the appendix we describe the extension to non-zero total momentum in the finite-volume frame.

\section{Review of the continuum derivation}
\label{sec:review_cont_deriv}

In this section, we outline the derivation for the $2\to2$ quantization condition and indicate which steps will be modified by the inclusion of $a^2$-effects. The derivation given below is based on refs.~\cite{Luscher:1986pf,Kim:2005gf,Hansen:2012tf,Briceno:2014oea} and the presentation closely follows refs.~\cite{Kim:2005gf,Hansen:2019nir}. The aforementioned references use a mix of Euclidean and Minkowski signatures, which is immaterial since the continuum finite-volume energy levels are metric-independent. However, discretized Minkowski and discretized Euclidean formulations can generically lead to different $a^2$ effects. Since the numerical lattice QCD calculations that we aim to describe are necessarily Euclidean, we will use this formulation throughout.

\subsection{The scattering amplitude and Bethe-Salpeter kernel}

\begin{figure}
\centering
\includegraphics{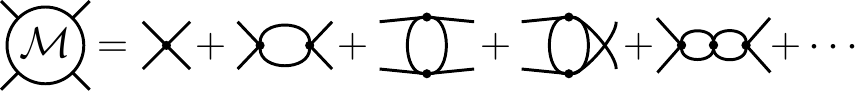}
\caption{The diagrams contributing the infinite-volume scattering amplitude in a theory containing a four-point vertex. External legs are amputated, and external momenta are placed on shell.}
\label{fig:scattering_amp_diag_def}
\end{figure}

The physical observable we aim to extract is the infinite-volume $2\to2$ scattering amplitude for two identical spin-zero particles with mass $m$. This is denoted by $\mathcal{M}({k}_{\sf in},{k}_{\sf out})$ and is shown diagrammatically in figure~\ref{fig:scattering_amp_diag_def}. Here ${k}_{\sf in}$ refers to one of the two incoming four-momenta and ${k}_{\sf out}$ to one of the two outgoing. As we are using Euclidean signature, these satisfy $k^2 = k_4^2 + \vec k^2$.

Throughout the main text, we assume vanishing spatial momentum such that the total incoming four-momentum is given by
\begin{equation}
P = (P_4, \vec{0}) \,. \label{eq:total_mom}
\end{equation}
The value of $P$ then fixes the other incoming and outgoing momenta as $P-k_{\sf in} $ and $P-k_{\sf out} $, respectively.
By analytically continuing $k_{{\sf in},4}$, $k_{{\sf out},4}$, and $P_4$ to be imaginary, one can define the physical, on-shell scattering amplitude. Here it is convenient to define $P_4 = i E$, where $E$ is simultaneously the finite-volume-frame energy and the center-of-mass frame (CMF) energy of the system. We require the CMF energy to be below the lowest multi-particle threshold, $E_{\sf{thresh}}$, and above the highest left-hand branch point, $E_{\sf{cut}}$: $ E_{\sf{cut}} < E < E_{\sf{thresh}}$, such that only the two-particle states are kinematically accessible. In the case where the upper limit is a three-particle threshold, significant progress has been made in extending the range of validity. See, for example, refs.~\cite{\ThreeBodyPapers}. Similarly, see refs.~\cite{\LeftHandCut} for recent discussion on how to treat energies below the lower limit.

As the infinite-volume, continuum system is $O(4)$ invariant, there is no loss of generality in choosing the CMF. However, in the finite-volume case, the boundary conditions define a frame and, for this reason, assuming vanishing momentum in this frame does lead to a loss of generality. We give the extension of the quantization condition incorporating discretization effects with non-zero total spatial momentum in appendix~\ref{app:non_zero_mom}.

In this work we will also consider quantities such as $\mathcal M(k_{\sf in},k_{\sf out})$ with off-shell momenta. This is only meaningful in the context of a particular operator used to construct the scattering amplitude. Specifically, we define the four-point correlator as
\begin{equation}
C(k_{\sf in},k_{\sf out})=\int \! d^4x \int \! d^4y \int \! d^4x' \, e^{i k_{\sf in} \cdot x + i (P-k_{\sf in}) \cdot y-i k_{\sf out} \cdot x'}\langle 0|{\text T} \! \left[\phi(x)\phi(y)\phi(x')\phi(0)\right]|0\rangle \,,
\end{equation}
and the fully dressed propagator as
\begin{equation}
D(k)=\int \! d^4x \, e^{i k \cdot x}\langle 0|{\text T} \! \left[\phi(x)\phi(0)\right]|0\rangle = \frac{Z(k)}{k^2 + m^2} \,,
\end{equation}
where the last step defines a particular off-shell continuation of the wavefunction renormalization $Z(k)$. The next step is to use amputation and inclusion of $Z(k)$ to define
\begin{multline}
\label{eq:amp_def}
\mathcal{M}(k_{\sf in},k_{\sf out}) = \sqrt{Z(k_{\sf in})Z(P-k_{\sf in})Z(k_{\sf out})Z(P-k_{\sf out})} \\
\times D(k_{\sf in})^{-1} D(P-k_{\sf in})^{-1} D(P-k_{\sf out})^{-1} D(k_{\sf out})^{-1} \, C(k_{\sf in},k_{\sf out}) \,.
\end{multline}
When external legs are placed on-shell
\begin{align}
k_{4, \sf{in}} & =i\omega_{k_{\sf in}}=i\sqrt{\vec{k}_{\sf in}^2+m^2}\,, \\
k_{4, \sf{out}} & =i\omega_{k_{\sf out}}=i\sqrt{\vec{k}_{\sf out}^2+m^2}\,,
\end{align}
then this definition agrees with the scattering amplitude, following the Lehmann-Symanzik-Zimmermann (LSZ) reduction formula. This also requires enforcing the on-shell condition for the remaining two particles, with momenta $P-k_{\sf in}$ and $P-k_{\sf out}$. The latter is achieved by fixing $E=2\omega_{k_{\sf in}}=2\omega_{k_{\sf out}}$, where the second equality can be phrased as $\vert \vec{k}_{\sf in} \vert = \vert \vec{k}_{\sf out} \vert$.

The infinite-volume scattering amplitude obeys the following integral equation:
\begin{align}
\label{eq:M_integral}
\mathcal{M}(p_{\sf in},p_{\sf out})=B(p_{\sf in},p_{\sf out})+B(p_{\sf in})\otimes \mathcal{I}\otimes \mathcal{M}(p_{\sf out}) \,,
\end{align}
where we have switched to $p$ as we will use $k$ to to denote an internal integrated momentum in the following.

\begin{figure}
\centering
\includegraphics{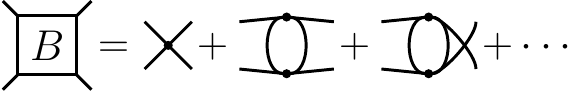}
\caption{The diagrams contributing to the infinite-volume BS kernel in a theory containing a four-point vertex. As with the scattering amplitude, external legs are amputated and replaced with the square-root of the single-particle residue.}
\label{fig:bs_kernel_def}
\end{figure}

We have also introduced the following:
\begin{itemize}
\item $B(p_{\sf in},p_{\sf out})$ is the two-particle Bethe-Salpeter (BS) kernel, which consists of all $2\to2$ diagrams that are two-particle irreducible in the channel carrying total momentum $P$, sometimes called the $s$-channel. In other words, $B$ contains all diagrams except those which fall apart when any two internal propagators carrying total momentum $P$ are cut. This is sketched diagrammatically in figure~\ref{fig:bs_kernel_def}. One can also regard eq.~\eqref{eq:M_integral} as the definition of the BS kernel, since all other quantities are defined independently.
\item $\otimes \mathcal{I}\otimes$ is an operator that acts by integrating over the loop momentum of the two-particle $s$-channel loop. Placed between two generic functions of the loop momenta $k$, denoted $L(k)$ and $R(k)$, it acts as:
\begin{align}
\label{eq:LR}
\begin{split}
L\otimes \mathcal{I} \otimes R&=\frac{1}{2}\int \frac{d^4k}{(2\pi)^4} L(k)\frac{D(k)}{Z(k)}\frac{D(P-k)}{Z(P-k)}R(k) \,, \\
&=\frac{1}{2}\int \frac{d^4k}{(2\pi)^4} L(k)\frac{1}{k^2+m^2}\frac{1}{(P-k)^2+m^2}R(k) \,.
\end{split}
\end{align}
Note that the integrated momentum is not written as an argument on the left-hand side.
In the derivation below, $L$ and $R$ will be replaced by either BS kernels or the scattering amplitudes, both of which include two momentum coordinates. As in eq.~\eqref{eq:M_integral}, whenever one of these arguments is suppressed, it is understood that it corresponds to an integrated coordinate. Note that, although the fully dressed propagators are included in the loop, these are rescaled by the explicit factors of $Z(k)$ and $Z(P-k)$ in eq.~\eqref{eq:LR}, effectively leading to non-interacting propagators. This follows from the definitions of the BS kernel and scattering amplitude, which include relative factors of $Z(k)$ as compared to the correlation functions, as in eq.~\eqref{eq:amp_def}.
\end{itemize}

\subsection{Finite-volume set-up}

Turning now to the finite-volume system, we take the temporal direction to be infinite and the three spatial directions to be finite with periodicity $L$. A consequence of applying periodic boundary conditions to the underlying fields is that the spatial momenta are quantized as integer-vector multiples of $(2\pi/L)$: $\vec k \in (2 \pi/L) \mathbb Z^3$\,.

The interacting finite-volume energy levels, which we aim to relate to the physical scattering amplitude, manifest as poles on the imaginary axis of any finite-volume Euclidean correlation function. One viable correlator containing these poles is a finite-volume version of $\mathcal M$, denoted by $\mathcal{M}_L$. Below, we will refer to this somewhat loosely as the finite-volume scattering amplitude. We emphasize, however, that this quantity does not represent a scattering process and is only relevant for the positions of its poles.

$\mathcal M_L$ is defined diagrammatically by following the expansion of the infinite-volume scattering amplitude and replacing all integrals over spatial momenta with sums:
\begin{align}
\underbrace{\, \int \! \frac{d^3\vec{k}}{(2\pi)^3}}_{\text{diagrams in}\ \mathcal{M}}\quad \mapsto\quad \underbrace{\frac{1}{L^3}\sum_{\vec{k} \in (2 \pi/L) \mathbb{Z}^3}.}_{\text{diagrams in}\ \mathcal{M}_L}
\end{align}

The authors of refs.~\cite{Luscher:1986pf,Kim:2005gf} have shown that, for any set of loops in which all particles cannot be simultaneously placed on-shell, the finite-volume sums and the infinite-volume integrals give the same value, up to exponentially suppressed corrections that scale as $e^{-mL}$, where $m$ is the smallest mass scale of the theory. As is standard in finite-volume scattering relations, we ignore such exponentially suppressed corrections in this work.

To understand the connection between on-shell intermediate states and the nature of $L$ scaling, one starts with the Poisson summation formula:
\begin{equation}
\label{eq:poisson}
\frac{1}{L^3} \sum_{\boldsymbol k \in (2 \pi/L) \mathbb{Z}^3}f(\boldsymbol k)=\sum_{\boldsymbol n\in \mathbb{Z}^3}\int \! \frac{d^3 \boldsymbol k}{(2\pi)^3} e^{-i \boldsymbol n \cdot \boldsymbol k L}f(\boldsymbol k) \,.
\end{equation}
This implies that the sum on the left-hand side and the integral ($(\boldsymbol n = \boldsymbol 0)$ term on the right-hand side) differ by exponentially suppressed scaling whenever $f(\boldsymbol k)$ has a domain of analyticity such that the $\vert \boldsymbol k \vert$ integration can be deformed into the complex plane to give a damping contribution in $ e^{-i \boldsymbol n \cdot \boldsymbol k L}$.

For a generic perturbative diagram, this analytic continuation is restricted when poles occur, and these exactly correspond to on-shell intermediate states. With a total CMF energy restricted as $E_{\sf{cut}}<E<E_{\sf{thresh}}$, on-shell contributions only arise from two-particle $s$-channel loops (in which two particles carry the total momentum $P$). Therefore, up to exponentially suppressed corrections, all momentum sums can be replaced with integrals, except when the sum runs over the momentum within such a two-particle loop.

It follows that, for both the fully dressed single-particle propagator and the BS kernel, one can replace the finite-volume version with its infinite-volume counterpart. The finite-volume correlator $\mathcal M_L$ then satisfies the equation:
\begin{align}
\label{eq:int_eq_finite}
\mathcal{M}_L(p_{\sf in},p_{\sf out})=B(p_{\sf in},p_{\sf out})+B(p_{\sf in})\otimes \mathcal{S}\otimes \mathcal{M}_L(p_{\sf out}) \,.
\end{align}
The new operator $\otimes S \otimes$ is analogous to $\otimes \mathcal{I}\otimes$, but with the integral replaced by a sum
\begin{align}
L\otimes \mathcal S \otimes R \equiv \frac{1}{2}\int \! \frac{d k_4}{2\pi}\frac{1}{L^3}\sum_{\vec{k}\in (2 \pi/L) \mathbb{Z}^3} L(k)\frac{D(k)}{Z(k)}\frac{D(P-k)}{Z(P-k)}R(k) \,.
\end{align}

\subsection{Sum-integral difference}

As discussed above, $\otimes\mathcal{S}\otimes$ and $\otimes\mathcal{I}\otimes$ represent sums or integrals over two-particle $s$-channel loops in which the particles can go on-shell. To relate the finite- and infinite-volume scattering amplitudes, we make the following replacement:
\begin{align}
\label{eq:f_insert}
L\otimes\mathcal{S}\otimes R=L\otimes\mathcal{I}\otimes R+L\boldsymbol{F}R \,,
\end{align}
where $\boldsymbol{F}$ is defined as the difference
\begin{align}
\label{eq:sum_int_def}
\begin{split}
L\boldsymbol{F}R&=\frac{1}{2}\int \frac{dk_4}{2\pi}\left[\frac{1}{L^3}\sum_{\vec k}-\int \frac{d^3 \vec{k}}{(2\pi)^3}\right] L(k)\frac{1}{k^2+m^2}\frac{1}{(P-k)^2+m^2}R(k) \,, \\
&=-\frac{1}{2}\left[\frac{1}{L^3}\sum_{\vec k}-\int \frac{d^3 \vec{k}}{(2\pi)^3}\right] L(k_4=i\omega_k,\vec{k})\frac{1}{2\omega_k E}\frac{1}{E-2\omega_k-i\epsilon}R(k_4=i\omega_k,\vec{k}).
\end{split}
\end{align}
In the second line, we have computed the $k_4$ integral, closing the contour above. Only the pole at $k_4=i\omega_k=i\sqrt{\vec{k}^2+m^2}$ contributes a term with poles for real $\boldsymbol k$, and as a result, the other terms lead to exponentially suppressed contributions (following the discussion around eq.~\eqref{eq:poisson}) and are dropped. The $i \epsilon$ prescription in the second line arises from analytically continuing $P_4 \to i E$.

In evaluating the $k_4$ integral, one additionally requires that any contributions from singularities in $L(k) R(k)$ lead to smooth $\vert \boldsymbol k \vert$-dependence and thus to terms scaling as $e^{-mL}$. This holds because the functions will ultimately be replaced by BS kernels, which have this property, again due to the lack of on-shell intermediate states. Our next step is to rewrite eq.~\eqref{eq:sum_int_def} as
\begin{align}
L\boldsymbol{F}R=-\frac{1}{2}\left[\frac{1}{L^3}\sum_{\vec k}-\int \frac{d^3 \vec{k}}{(2\pi)^3}\right] {\sf L}(|\vec{k}|^2,\hat{\vec{k}})\frac{1}{2\omega_k E}\frac{1}{E-2\omega_k-i\epsilon}{\sf R}(|\vec{k}|^2,\hat{\vec{k}}) \,,
\end{align}
where we introduced the sans-serif font to represent new coordinate dependence
\begin{align}
\begin{split}
{\sf L}(|\vec{k}|^2,\hat{\vec{k}})&=L(k_4=i\omega_k,\vec{k}) \,,\\
{\sf R}(|\vec{k}|^2,\hat{\vec{k}})&=R(k_4=i\omega_k,\vec{k}) \,,
\end{split}
\end{align}
and have separated the magnitude of the momentum ($\vert \boldsymbol k \vert^2$) from the angular dependence $\hat{\vec{k}}=\vec{k}/|\vec{k}|$.

The final important result, demonstrated in ref.~\cite{Luscher:1986pf} is that, again up to corrections suppressed as $e^{-mL}$, the functions $L$ and $R$ can be evaluated at the pole location. That is, we can set $\vec{k}^2$ to the roots of $E=2\omega_k$. Introducing the quantity $p^2$ via
\begin{align}
E^2=4(p^2+m^2) \,,
\end{align}
the prescription is to set $\vec{k}^2=p^2$ such that the two functions only depend on direction through $\hat{\vec{k}}$.

A subtlety of this step is that spurious singularities near $\boldsymbol k^2 = 0$ will arise if we simply replace ${\sf L}(|\vec{k}|^2,\hat{\vec{k}})$ with ${\sf L}(p^2,\hat{\vec{k}})$. To resolve this we first decompose the functions onto spherical harmonics as:
\begin{align}
\begin{split}
{\sf L}(\boldsymbol k^2,\hat{\vec{k}})&= \sqrt{4\pi} \, Y_{lm}(\hat{\vec k}) \, |\vec{k}|^l \, {\sf L}_{lm}(\boldsymbol k^2) \,, \\
{\sf R}(\boldsymbol k^2,\hat{\vec{k}})&= \sqrt{4\pi} \, Y^*_{l'm'}(\hat{\vec k}) \, |\vec{k}|^{l'} \, {\sf R}_{l'm'}(\boldsymbol k^2) \,,
\end{split}
\end{align}
where we have also scaled out powers of $\vert \boldsymbol k \vert$ such that ${\sf R}_{lm}(\boldsymbol k^2) $ and ${\sf L}_{lm}(\boldsymbol k^2)$ both approach a non-zero constant as $\vert \boldsymbol k \vert \to 0$, rather than vanishing as $\vert \boldsymbol k \vert^l$.
The following replacement can then be made within $\boldsymbol F$:
\begin{align}
\begin{split}
{\sf L}( \vert \boldsymbol k \vert^2 ,\hat{\vec{k}}) \ \ &\longrightarrow \ \ \sqrt{4\pi} \, Y_{lm}(\hat{\vec k}) \, |\vec{k}|^l \, {\sf L}_{lm}(p^2) \,,\\
{\sf R}( \vert \boldsymbol k \vert^2 ,\hat{\vec{k}}) \ \ &\longrightarrow \ \ \sqrt{4\pi} \, Y^*_{l'm'}(\hat{\vec k}) \, |\vec{k}|^{l'} \, {\sf R}_{l'm'}(p^2) \,.
\end{split}
\end{align}

Note that the decomposed functions, ${\sf L}_{lm}(p^2)$ and ${\sf R}_{lm}(p^2)$, no longer depend on the loop momenta and hence can be pulled out of the sum-integral difference.
The result of these manipulations is that the term with the $\boldsymbol{F}$-insertion can be treated as a matrix product:
\begin{align} L\boldsymbol{F}R&={\sf L}_{lm}(p^2)\boldsymbol{F}_{lm,l'm'}(E,L){\sf
R}(p^2)_{l'm'} \,,
\end{align}
where
\begin{align}
\label{eq:f_insert_def}
\boldsymbol{F}_{lm,l'm'}(E,L)&=-\frac{1}{2E^2}\left[\frac{1}{L^3}\sum_{\vec k}-\int \frac{d^3 \vec{k}}{(2\pi)^3}\right]\frac{4\pi Y_{lm}(\hat{\vec k})Y^*_{l'm'}(\hat{\vec k}) |\vec{k}|^{l+l'}}{(E-2\omega_k-i\epsilon)} \,.
\end{align}

It is this process of placing the loop momentum on-shell, decomposing the $L$ and $R$ functions onto spherical harmonics and forming the resultant $\boldsymbol{F}$ matrix that will be altered when including discretization effects. In particular, $a^2$-effects introduce angular dependence to the pole position, and this will make the subsequent decomposition of the $L$ and $R$ functions more complicated. In addition, the lattice artifacts will change the form of $\boldsymbol{F}$ and complicate the on-shell definition of adjacent quantities. The overall strategy of the derivation will otherwise be unchanged.

\subsection{Quantization condition}

Having defined $\boldsymbol F(E,L)$ and explained the substitution rule for relating sums and integrals, given by eq.~\eqref{eq:f_insert}, we are now in position to derive the quantization condition.

We first repeatedly substitute the left-hand side of eq.~\eqref{eq:M_integral} into the right-most factor on the right-hand side, to reach an infinite series for the scattering amplitude
\begin{multline}
\label{eq:infinite_M_expand}
\mathcal{M}(p_{\sf in},p_{{\rm out}})=B(p_{\sf in},p_{\sf out})+B(p_{\sf in})\otimes\mathcal{I}\otimes B(p_{\sf out}) \\ +B(p_{\sf in})\otimes\mathcal{I}\otimes B\otimes\mathcal{I}\otimes B(p_{\sf out})+\dots \,.
\end{multline}
The same procedure on eq.~\eqref{eq:int_eq_finite} yields
\begin{multline}
\label{eq:finite_M_expand}
\mathcal{M}_L(p_{\sf in},p_{\sf out})=B(p_{\sf in},p_{\sf out})+B(p_{\sf in})\otimes\mathcal{S}\otimes B(p_{\sf out}) \\ +B(p_{\sf in})\otimes\mathcal{S}\otimes B\otimes\mathcal{S}\otimes B(p_{\sf out})+\dots \,.
\end{multline}

Then applying eq.~\eqref{eq:f_insert} for all instances of $\mathcal S$, we derive the desired form for $\mathcal M_L(E)$:
\begin{align}
\label{eq:exp_cont}
\mathcal{M}_L(E)&=\sum_{n=0}^\infty B \left(\left[\otimes \mathcal{I}\otimes +\boldsymbol{F}\right] B\right)^n \,, \\
&=\sum_n\mathcal{M}\left[\boldsymbol{F}\mathcal{M}\right]^n \,, \\&=\frac{1}{\mathcal{M}^{-1}(E)-\boldsymbol{F}(E, L)} \,,
\end{align}
where each factor on the second and third line is understood as a matrix with joint $\{l,m\}$ indices. In the second line, various $\boldsymbol F$-independent series have been identified with eq.~\eqref{eq:infinite_M_expand} so that the dependence on the BS kernel can be completely removed, in favor of a dependence on the scattering amplitude. In order to give a matrix equation, we have also decomposed the external momenta of $\mathcal M_L(E)$ using spherical harmonics. Further details of this procedure can be found in refs.~\cite{Luscher:1986pf,Kim:2005gf,Hansen:2019nir}.

We deduce that, up to neglected $e^{- mL}$ and assuming $E_{\sf{cut}}<E<E_{\sf{thresh}}$, the finite-volume energy levels are given by the poles of $\mathcal M_L(E)$, which occur whenever:
\begin{align}
\label{eq:final_continuum}
\underset{lm}{\text{det}}\left[\mathcal{M}(E)^{-1}-\boldsymbol{F}(E, L)\right] \bigg \vert_{E=E_n(L)}=0 \,.
\end{align}
This completes our review of the quantization condition in the continuum. We have not given an explict discussion of $\mathcal M_{l m, l'm'}(E)$, a diagonal matrix with entries corresponding to partial-wave projected amplitudes. See also refs.~\cite{Luscher:1986pf,Kim:2005gf,Hansen:2019nir} and the discussion in section~\ref{sec:final_result}.

\section{Discretized quantization condition}
\label{sec:disc_deriv}

We now turn to the consequences of incorporating discretization effects in the derivation. These are included via improved Symanzik Effective Theory (SET)~\cite{Symanzik:1983dc,Bar:2004xp}, which encodes lattice artifacts into a continuum theory via irrelevant operators with coefficients that are proportional to $a^2$. The next subsection provides a review of the basics of SET. Importantly, we will see that the new operators can break $O(3)$ rotational symmetry.

All quantities introduced above will be revisited in the SET. For example, $B$, $\mathcal{M}$, and $D$, will be replaced by quantities denoted $B_a$, $\mathcal{M}_a$, and $D_a$, which refer to the infinite-volume BS kernel, scattering amplitude, and fully dressed propagator respectively, but now including additional contributions from the $O(a^2)$ vertices.

As in the continuum derivation, the notion of on-shell projection will play an important role here. The on-shell condition depends on the location of the single-particle poles, and this motivates our analysis of the single-particle propagator in sections~\ref{sec:simp_prop} and \ref{sec:disc_f}. In section~\ref{sec:disc_sum_int_diff}, we show how these effects are included in the sum-integral difference, leading to a modified $\boldsymbol{F}$ function, defined in eq.~\eqref{eq:f_disc_final}. Finally, in section~\ref{sec:final_result}, we combine these results to derive our main result, eq.~\eqref{eq:final_continuum}.

In the following, we will see that two key modifications arise in the manipulations analogous to those that tooks us from eq.~\eqref{eq:sum_int_def} to eq.~\eqref{eq:f_insert_def} above. First, additional $a$-dependent terms arise directly in the BS kernel and thus in the scattering amplitude from the extra vertices. Second, and of greater technical relevance, the step of placing internal legs on-shell is complicated as the pole position inherits angular dependence from $k_x^4+k_y^4+k_z^4$ contributions. This ultimately requires an additional decomposition in spherical harmonics, resulting in more indices on the finite-volume function.

Throughout the derivation, we only keep $O(a^2)$ effects while dropping $O(a^3)$ as well as possible logarithmic dependence.

\subsection{Review of Symanzik Effective Theory (SET)}

To incorporate the effects of discretized spacetime in QCD, the first step is to recognize that lattice QCD can be expressed as continuum QCD together with an infinite set of irrelevant operators, with coefficients suppressed by powers of the lattice spacing $a$. The SET~\cite{Symanzik:1983dc} is a systematic way to organize these operators, constrained only by the underlying symmetries of the lattice theory.

In a second step, the SET Lagrangian is matched to that of a low-energy effective field theory (EFT), leading to an infinite set of irrelevant (in the Wilsonian sense) terms, built form the low-energy degrees of freedom, again suppressed by powers of the lattice spacing. Assuming that some form of Symanzik improvement has been applied, such that the leading order corrections are of order $a^2$, then the EFT Lagrangian for this theory takes the form
\begin{align}
\label{eq:Lagrangian_discrete}
\mathcal{L}(x) =\mathcal{L}_{\sf cont}(x) +a^2 \mathcal{L}_{\sf disc}(x) +O(a^3) \,.
\end{align}
The lattice spacing breaks rotational symmetry and so the discretized Lagrangian will include terms such as
\begin{equation}
\mathcal{L}_{\sf disc}(x) \supset \sum_{\mu} \phi(x) \partial_\mu^4 \phi(x) \,,
\label{eq:disc_lagrangian_d4}
\end{equation}
leading to vertices depending on non-rotationally invariant combinations of momenta. In particular, we define
$[p]^4=\sum_\mu (p_\mu)^4$.

The derivation below assumes that no additional flavors or tastes are introduced through discretization and is therefore not applicable to staggered fermions, irrespective of whether the fermion determinant is rooted in the implementation. The application to improved Wilson, domain-wall and overlap fermions is immediate. The question as to whether the derivation applies to twisted mass fermions is subtle and future work is needed to investigate this context.

To explicitly see the relationship between the EFT and the full discretized theory, consider the free lattice propagator for a massless spin-zero particle:
\begin{align}
\Delta(k)= \left [ \left (\frac{2}{a} \right )^{\!2} \sum_\mu\sin^2 \left (\frac{k_\mu a}{2} \right ) \right ]^{-1},
\end{align}
where $\Delta(k)$
indicates that this is a bare propagator and the sum runs over the four spacetime dimensions.
For $ka\ll 1$, this can be expanded in powers of $a$ to give
\begin{align}
\label{eq:delta_expansion}
\Delta(k)=\frac{1}{k^2-a^2 \sum_\mu (k_\mu)^4/12} + O(a^4) =\frac{1}{k^2}+ \frac{1}{k^2} \frac{a^2\sum_\mu (k_\mu)^4}{12} \frac{1}{k^2} + O(a^4) \,.
\end{align}
We observe that the expanded form can be recovered via an EFT that includes a two-point vertex of the form given in eq.~\eqref{eq:disc_lagrangian_d4}.

From this example, we can see that the SET is only valid in the regime when all energy and momentum scales are much less than $1/a$. Importantly for our purposes, $\Delta(k)$ is periodic in each $k_\mu$ and it is convenient to restrict the latter to the first Brillouin zone $k_\mu\in (-\pi/a,\pi/a]$. In this region, the only pole is at $k=0$. However, the first expanded form (middle expression above) is no longer periodic and contains an additional pole on a manifold in $k$-space that includes, for example, the point $k=\left ( 0,0,0,\frac{2\sqrt 3}{a} \right )$.

This pole is an artifact of the approximation, which we expect to breakdown when $ka$ is order one or larger. To address this, we will introduce a smooth ultraviolet cutoff function in our derivation that removes unphysical poles. The most straightforward approach is to follow the three-particle formalism of refs.~\cite{\ThreeParticleCutoff} and use a cutoff that identically vanishes above a given energy threshold. Such a function is necessarily non-analytic and as a result, the neglected final-volume effects in the resulting quantization condition are not exponentially suppressed, but instead fall faster than any power of $1/L$.

As a final comment, we emphasize that any discussion of $a$ dependence is only meaningful once a scaling trajectory has been defined. Typically, in addition to defining all details of the lattice action, this requires selecting a set of parameters that are tuned to have their desired values at all $a$. For a 2+1 flavor lattice QCD calculation, for example, one can define the bare light and strange quark masses such that $m_\pi/m_\Omega$ and $m_K/m_\Omega$ have constant values for a range of lattice spacings. Then, by interpreting the calculations to be at fixed $m_\Omega$ and setting a value for the latter, for example from the Particle Data Group~\cite{ParticleDataGroup:2024cfk}, one infers a relation between the bare gauge coupling, $\beta$, and the lattice spacing in physical units. In this way all other quantities can be expressed as a function of $a$ for fixed $m_\pi/m_\Omega$, $m_K/m_\Omega$ and $m_\Omega$, or any analogous set of quantities.

In the following we will generically refer to the scattering hadron as a pion with mass $m_\pi$, but the results are applicable to any identical two-particle state. If a trajectory is chosen for which $m_\pi$ is not independent of $a$, then one arrives at a function $m_\pi(a)$. We find it convenient to also define
\begin{equation}
m_{\sf c} = \lim_{a \to 0} m_\pi(a) \,,
\end{equation}
where the subscript ${\sf c}$ stands for continuum.
One can then write
\begin{align}
m_\pi^2=m_{\sf c}^2 + a^2 \left [ \frac{\partial m_\pi(a)^2}{\partial a^2}\bigg \vert_{a = 0} \right ] +O(a^3, a^2 \log(a)) \,,
\end{align}
where $m_\pi$ without an argument will always refer to the lattice pion mass.

\subsection{Simplifying the Propagator}
\label{sec:simp_prop}

To derive the discretized quantization condition, we first need a useful expression for the single-hadron propagator. The final result of this subsection is a simplified form for the propagator, that we use to define a modified version of the L{\"u}scher finite-volume function.

In a theory with discretization effects encoded at $O(a^2)$ via SET, the most general form for a fully dressed single-hadron propagator is
\begin{align}
\label{eq:general_prop_dup1}
D_a(k)=\int d^4x \, e^{-ikx} \, \<\phi(x)\phi^\dagger(0)\>_a =\frac{z_a(k)}{k^2+m^2_{\sf c}+a^2 f(k)} \,,
\end{align}
where $m_{\sf c}$ is the pole position in the continuum limit. There is some freedom here in the division between $z_a(k)$ and $f(k)$. One viable definition is to require $z_a(k) = Z(k)$, i.e.~take the numerator function to match the function defined in the continuum. This then fixes the definition of $f(k)$. Alternatively, one can include some $a^2$ effects in the definition of $z_a(k)$ and modify $f(k)$ appropriately.

The next step is to restrict attention to the on-shell function $f(i\omega_{{\sf c}, k},\vec{k})$, where $\omega_{{\sf c}, k} = \sqrt{\boldsymbol k^2 + m_{\sf c}^2}$ can be used, since the difference between this and any discretized version is beyond the order we control. With this quantity in hand, we write
\begin{equation}
D_a(k)= \frac{Z_a(k)}{k_4^2+\vec{k}^2+m^2_{\sf c}+a^2 f(i\omega_{{\sf c}, k},\vec{k})}
+ {O}(a^4) \,,
\end{equation}
where we have introduced
\begin{align}
Z_a(k)= z_a(k) \left [ 1+a^2 \frac{f(k_4,\vec{k})-f(i\omega_{{\sf c}, k},\vec{k})}{k^2+m^2_{\sf c}}\right ]^{-1} \,.
\end{align}

Next we introduce ${\sf f}(\boldsymbol k) = f(i\omega_{{\sf c}, k},\vec{k})-f(i\omega_0,\vec{0})$ and define a lattice version of the single-particle energy
\begin{align}
\label{eq:disc_omega}
\Omega_{k}^2 & \equiv \vec{k}^2+m^2_{\sf c}+a^2 f(i\omega_{{\sf c}, k},\vec{k}) = \vec{k}^2+m^2_{\sf \pi}+a^2 \f (\vec{k}) \,.
\end{align}
The last equality, in terms of $m_\pi^2$ and $\f(\boldsymbol k)$, is the most convenient version for the subsequent steps of the derivation.

The final step is to expand the numerator function about the pole position. The leading order contribution to such an expansion is given by defining
\begin{align}
\label{eq:residue}
\Z (\vec{k}) \equiv \lim_{k_4\to i \omega_{{\sf c}, k}} Z_a(k) \,.
\end{align}
We thus reach the main result of this section, that the lattice propagator near the pole can be written as
\begin{align}
\label{eq:prop_final_latt}
D_a(k)=\frac{\Z(\vec{k})}{k_4^2+ \boldsymbol k^2 + m_\pi^2 + a^2 {\sf f}(\boldsymbol k)} + O \left (a^3, (k_4^2-\omega_{{\sf c}, k}^2)^0 \right ) \,.
\end{align}

\subsection{Determining \texorpdfstring{$\f(\boldsymbol k)$}{the f-function}}
\label{sec:disc_f}

In the steps below we will see that $\f(\boldsymbol k)$ plays an important role in our final result, while ${\sf Z}_a(\boldsymbol k)$ does not. Therefore, here we focus on the extraction of the former. From dimensional and symmetry considerations, $\f(\boldsymbol k)$ must take the form
\begin{equation}
\f(\boldsymbol k) = \alpha_1 (\boldsymbol k^2)^2 + \alpha_2 [\boldsymbol k]^4 + \alpha_3 \boldsymbol k^2 m_\pi^2 + O (a \mu^5) \,,
\end{equation}
where $\mu$ here stands for a generic quantity with energy dimension one. This simple expression follows from the fact that the lattice spacing is encoded via couplings in the SET, that dimensional consistency dictates the powers of momentum and mass that can enter with these couplings, and that we are dropping higher order powers in $a$ throughout. Finally we note that logarithmic corrections to $O(a^2)$ scaling have recently been investigated in refs.~\cite{Husung:2021mfl,Husung:2022kvi}. We do not explicitly include such effects here but note that they can be incorporated into the $a$-dependence of $m_\pi(a)$ as well as a potential lograthimic dependence in the $\alpha_i$ couplings.

Fourier transforming eq.~\eqref{eq:prop_final_latt} to the time-momentum representation, we define
\begin{align}
G_a(\tau, \boldsymbol k) = \int \frac{d k_4}{2 \pi} e^{i k_4 \tau} D_a(k) = \frac{\Z(\boldsymbol k)}{2 \Omega_k} e^{-\Omega_k \vert \tau \vert} + \sum_{n = 1}^\infty c_n(L) e^{-E_n(L) \vert \tau \vert} + O (a^3 ) \,.
\end{align}
The second term represents multi-hadron excited states arising from the $O((k_4-\Omega_k)^0)$ terms, and we will ignore these in the following. The first term is the single-hadron contribution. By evaluating this for various $\boldsymbol k$, one can, in principle, overconstrain the coefficients $\alpha_i$ and determine them. To this end we stress that the $(\boldsymbol k^2)^2 $ and $ [\boldsymbol k]^4$ terms can only be separated by momenta for which $\boldsymbol k^2$ are equivalent but $[\boldsymbol k]^4$ are not. The lowest such momenta in a periodic cubic box are $\boldsymbol k_{[003]} = (2 \pi / L) (0,0,3)$ and $\boldsymbol k_{[122]} = (2 \pi / L) (1,2,2)$. Denoting the single-hadron lattice energy evaluated at this momentum by $\Omega_{[n_xn_yn_z]}$, one finds
\begin{align}
\Omega_{[122]}^2-\Omega_{[003]}^2 = a^2 [ \f(\boldsymbol k_{[122]})-\f(\boldsymbol k_{[003]}) ] + O (\mu a^3) =-a^2 \alpha_2 \frac{4 \pi^2}{L^2} 4 8 + O (\mu a^3) \,.
\end{align}
Interestingly, for a typical lattice QCD geometry of $L/a = 48$ the term on the right-hand side takes on the numerical value of $-0.82 \alpha_2$ meaning that the coefficient is order one.

The feasibility of determining these coefficients in practice will depend greatly on the details of the calculation. In the following we take the coefficients as known and consider the consequences of this for the quantization condition.

\subsection{Sum-integral difference with lattice artifacts}

\label{sec:disc_sum_int_diff}

In this section we use the propagator given in eq.~\eqref{eq:prop_final_latt} to rederive the results from eqs.~\eqref{eq:sum_int_def} to~\eqref{eq:f_insert_def} for the discretized case. The final form for the finite-volume function with $a$-dependence, which we call $\boldsymbol{F}_a$, is given in eq.~\eqref{eq:f_disc_final}, and the recipe for placing the adjacent functions on-shell and decomposing onto spherical harmonics is given in the text preceding that equation.

Recalling that we set $P=(P_4, \boldsymbol 0)$ until appendix~\ref{app:non_zero_mom}, the sum-integral difference of the $s$-channel loop between two arbitrary analytic functions $L_a$ and $R_a$ is given by:
\begin{align}
\label{eq:LFR_disc}
L_a\boldsymbol{F}_{\!a} R_a&=\frac{1}{2}\int \frac{dk_4}{2\pi}\left[\frac{1}{L^3}\sum_{\vec k}-\int \frac{d^3 \vec{k}}{(2\pi)^3}\right] L_a(k)\frac{{\sf H}(\vec{k})}{k_4^2+\Omega_k^2}\frac{1}{(P_4-k_4)^2+\Omega_k^2}R_a(k) \,,
\end{align}
where ${\sf H}(\vec{k})$ is a smooth cutoff that is 1 for a region of $\vert \boldsymbol k \vert$ of width larger than or equal to $m$, including the physical pole position, and then smoothly interpolates to 0 for $m \ll \vert \boldsymbol k \vert \ll 1/a$ with the requirement that it is identically zero in a region of $\boldsymbol k$ where an unphysical pole might arise, so that the latter does not appear in $\boldsymbol{F}_{\!a}$.

An example of a function that satisfies these requirements is ${\sf H}(\boldsymbol{k}) = J(x(\boldsymbol k, \Lambda, \Gamma))$ where
\begin{equation}
x(\boldsymbol k, \Lambda, \Gamma) = \frac{\Lambda^2-\boldsymbol k^2}{\Gamma^2} \,,
\label{eq:xdef}
\end{equation}
and
where $J(x)$ is the function introduced in ref.~\cite{Hansen:2014eka}, which we repeat here for convenience
\begin{align}
J(x)=\begin{cases} 0 & x<0\\ \text{exp}\left(-\frac{1}{x}\text{exp}[-\frac{1}{1-x}]\right) & 0<x<1\\ 1& x>1 \end{cases} \,.
\label{eq:Jdef}
\end{align}
Here $\Lambda$ and $\Gamma$ are free parameters, that can be chosen to define the regions in which the cutoff is equal to 1 and zero.

Evaluating the $k_4$ integral, analytically continuing $P_4$ to be along the positive imaginary axis and substituting $P_4 = i E$, we find
\begin{align}
L_a\boldsymbol{F}_{\!a} R_a&=-\frac{1}{2}\left[\frac{1}{L^3}\sum_{\vec k}-\int \! \frac{d^3 \vec{k}}{(2\pi)^3}\right] {\sf L}_a(\vec{k}) \frac{1}{2\Omega_k E}\frac{{\sf H}(\vec{k})}{E-2\Omega_k + i \epsilon} {\sf R_a}(\vec{k}) \,,
\end{align}
where we have again introduced the sans-serif font to denote quantities that have been evaluated on-shell:
\begin{align}
{\sf L}_a(\vec{k}) & \equiv L_a(k_4=i\Omega_k,\vec{k}) \,, \qquad {\sf R}_a(\vec{k}) \equiv R_a(k_4=i\Omega_k,\vec{k}) \,.
\end{align}
As with eq.~\eqref{eq:LR} above, we stress that a rescaling by ${\sf Z}_a(\boldsymbol k)$ is included in the definition of ${\sf L}_a$ and ${\sf R}_a$. This will ultimately affect the definition of $\mathcal M_a$, the discretized scattering amplitude extracted from our formalism, as is explained in detail in section~\ref{sec:final_result}.

The next step is to multiply the numerator and denominator by $E + 2 \Omega_k$ to write
\begin{align}
L_a \boldsymbol{F}_{\!a} R_a =-\frac{1}{2} \left[\frac{1}{L^3}\sum_{\vec k}-\int \frac{d^3 \vec{k}}{(2\pi)^3}\right] {\sf L}(\boldsymbol k^2, \hat{\boldsymbol k})\frac{2\Omega_k+E}{2\Omega_k E}\frac{{\sf H}(\vec{k})}{E^2-4\Omega_k^2+ i \epsilon}{\sf R}(\boldsymbol k^2, \hat{\boldsymbol k}) \,.
\end{align}
The location of the pole is determined by the solutions of $E=2\Omega_k$ and we can evaluate $\Omega_k$ at the pole location to make the replacement
\begin{align}
\frac{2\Omega_k+E}{2\Omega_kE}\mapsto \frac{2}{E} \,,
\end{align}
which holds up to exponentially suppressed corrections.

Next, introducing the standard relation
\begin{equation}
p^2 = E^2/4-m_\pi^2 \,,
\end{equation}
we can write the sum-integral difference as
\begin{align}
\label{eq:LFR_before_decomp}
L_a \boldsymbol{F}_{\!a} R_a =-\frac{1}{4 E} \left[\frac{1}{L^3}\sum_{\vec k}-\int \frac{d^3 \vec{k}}{(2\pi)^3}\right] {\sf L}_a(\boldsymbol k^2, \hat{\boldsymbol k})\frac{{\sf H}(\vec{k})}{p^2-\vec{k}^2-a^2 \f (\vec{k}) + i \epsilon}{\sf R}_a(\boldsymbol k^2, \hat{\boldsymbol k}) \,.
\end{align}

At this stage we need to remove the directional dependence from the endcap factors. We define $[{\sf L}_a]_{l m}(\boldsymbol k^2)$ and $[{\sf R}_a]_{lm}(\boldsymbol k^2)$ via
\begin{align}
\begin{split}
\label{eq:first_decomp}
[{\sf L}_a](\boldsymbol k^2, \hat {\boldsymbol k}) & = [{\sf L}_a]_{lm}(\boldsymbol k^2) \sqrt{4 \pi} \vert \boldsymbol k \vert^l Y_{lm}(\hat {\boldsymbol k}) \,, \\ [{\sf R}_a](\boldsymbol k^2, \hat {\boldsymbol k}) & = [{\sf R}_a]_{lm}(\boldsymbol k^2) \sqrt{4 \pi} \vert \boldsymbol k \vert^l Y^*_{lm}(\hat {\boldsymbol k}) \,.
\end{split}
\end{align}
From here we can write
\begin{align}
L_a \boldsymbol{F}_{\!a} R_a =-\frac{1}{4 E} \left[\frac{1}{L^3}\sum_{\vec k}-\int \frac{d^3 \vec{k}}{(2\pi)^3}\right] [{\sf L}_a]_{lm}(\boldsymbol k^2)\frac{4 \pi \vert \boldsymbol k \vert^{l+l'} Y_{lm}(\hat {\boldsymbol k}) Y_{l'm'}(\hat {\boldsymbol k}) {\sf H}(\vec{k})}{p^2-\vec{k}^2-a^2 \f (\vec{k}) + i \epsilon}[{\sf R}_a]_{l'm'}(\boldsymbol k^2) \,.
\end{align}

We observe that, up to exponentially suppressed volume effects, one can replace $\boldsymbol k^2 \to p^2-a^2 {\sf f} (\boldsymbol k)$ in both of the partial wave project endcaps
\begin{align}
\label{eq:on_shell_replacement}
[{\sf L}_a]_{lm}(\boldsymbol k^2) \to [{\sf L}_a]_{lm}(p^2-a^2 {\sf f}(\boldsymbol k)) \,, \qquad [{\sf R}_a]_{l'm'}(\boldsymbol k^2) \to [{\sf R}_a]_{l'm'}(p^2-a^2 {\sf f}(\boldsymbol k)) \,.
\end{align}
The resulting functions depend on $p^2$, $\boldsymbol k^2$ and $\hat {\boldsymbol k}$ and this motivates us to perform a second decomposition, to write
\begin{align}
\label{eq:second_decom_LR}
[{\sf L}_a]_{lm}(p^2-a^2 {\sf f}(\boldsymbol k)) & = [{\sf L}_a]_{lm,\alpha \beta}(p^2, \boldsymbol k^2) \vert \boldsymbol k \vert^\alpha Y_{\alpha \beta}(\hat {\boldsymbol k}) \,, \\
[{\sf R}_a]_{l'm'}(p^2-a^2 {\sf f}(\boldsymbol k)) & = [{\sf R}_a]_{l'm',\alpha' \beta'}(p^2, \boldsymbol k^2) \vert \boldsymbol k \vert^{\alpha'} Y^*_{\alpha' \beta'}(\hat {\boldsymbol k}) \,,
\end{align}
where the directional dependence comes only through $\f(\boldsymbol k)$.

The coefficients on the right-hand side can be placed on-shell a second time using $\boldsymbol k^2 \to p^2$. At this level, the $a^2$ corrections to the pole position lead to overall $a^4$ and higher effects that are beyond the order we control. Adopting the shorthand $[{\sf L}_a]_{lm, \alpha \beta}(p^2, p^2) = [{\sf L}_a]_{lm, \alpha \beta}(p^2)$ and similar for the right endcap, we deduce
\begin{align}
L_a \boldsymbol{F}_{\!a} R_a = [{\sf L}_a]_{lm, \alpha \beta}(p^2) [\boldsymbol{F}_{\!a}]_{lm, \alpha \beta, l'm', \alpha' \beta'}(E, L) [{\sf R}_a]_{l'm', \alpha' \beta'}(p^2) \,,
\end{align}
where we have defined
\begin{multline}
[\boldsymbol{F}_{\!a}]_{lm, \alpha \beta, l'm', \alpha' \beta'}(E, L) \equiv-\frac{1}{4 E} \left[\frac{1}{L^3}\sum_{\vec k}-\int \frac{d^3 \vec{k}}{(2\pi)^3}\right]
\\ \times \frac{16 \pi^2 \vert \boldsymbol k \vert^{l+l'+\alpha+\alpha'} Y_{lm}(\hat {\boldsymbol k}) Y_{\alpha \beta}(\hat {\boldsymbol k}) Y^*_{l'm'}(\hat {\boldsymbol k}) Y^*_{\alpha' \beta'}(\hat {\boldsymbol k}) \, {\sf H}(\vec{k})}{p^2-\vec{k}^2-a^2 \f (\vec{k}) + i \epsilon} \,.
\label{eq:f_disc_final}
\end{multline}
This is the main result of this subsection.

A few comments are in order:
\begin{enumerate}
\item Only the $\alpha = \beta = 0$ components of $[{\sf L}_a]_{lm, \alpha \beta}(p^2)$ and $[{\sf R}_a]_{lm, \alpha \beta}(p^2)$ have a non-zero continuum limit. Thus, all non-trivial components in $\alpha, \beta, \alpha', \beta'$ are interpreted as mixing due to the lattice breaking of rotational symmetry (distinct from the mixing due to the finite-volume breaking of the symmetry, encoded in $l,m,l',m'$).
\item The $\boldsymbol{F}_{\!a}$ matrix and the endcap factors here are defined with rescaling of powers of the relevant momentum component. A consequence of this is that $[{\sf L}_a]_{lm, \alpha \beta}(p^2)$ and $[{\sf R}_a]_{lm, \alpha \beta}(p^2)$ are not suppressed by powers of $p^2$ but instead approach a constant for $p^2 \to 0$ in all components. The suppression is instead encoded in $\boldsymbol{F}_{\!a}$. It is, of course, possible to rescale $[{\sf L}_a]$, $[{\sf R}_a]$ and $\boldsymbol{F}_{\!a}$ to remove this feature, but we have chosen not to do so.
\end{enumerate}

\subsection{Discretized quantization condition}
\label{sec:final_result}

We now collect the results above to present the discretized quantization condition. Using the new $\boldsymbol{F}_{\!a}$ matrix, the expansion of eq.~\eqref{eq:exp_cont} can be carried through identically. We hence find that the finite-volume energy levels are given by the roots of the following determinant:
\begin{align}
\label{eq:final_QC}
\det_{lm, \alpha\beta} \left [\mathcal{M}_a(E)^{-1}-\boldsymbol{F}_{\!a}(E,L) \right ] = 0 \,,
\end{align}
where $\boldsymbol{F}_{\!a}$ is given in eq.~\eqref{eq:f_disc_final} and the indices are given as a reminder that the determinant is over two sets.

It is worth considering in detail the meaning of the new scattering amplitude, $\mathcal{M}_a(E)$. To explain this, we first construct the scattering amplitude as a function of the incoming and outgoing 4-momenta: $\mathcal{M}_a(k_{\sf in}, k_{\sf out})$. The definition is given diagrammatically as follows: First one takes all four-point diagrams built from vertices that survive the continuum limit, as well as vertices from the $O(a^2)$ part of the Lagrangian. Denote the four-point function constructed in this way through $O(a^2)$ by $C_a(k_{\sf in}, k_{\sf out})$. Next amputate the fully dressed external legs to reach $C^{\sf amp}_a(k_{\sf in}, k_{\sf out})$. Our discretized version of the scattering amplitude is then given by
\begin{equation}
\mathcal{M}_a(k_{\sf in}, k_{\sf out}) = {\sf Z}_a(\vec{k}_{\sf in}) {\sf Z}_a(\vec{k}_{\sf out}) C^{\sf amp}_a(k_{\sf in}, k_{\sf out}) \,,
\label{eq:adep_scattering_amp}
\end{equation}
where ${\sf Z}_a(\boldsymbol k)$ is defined in eq.~\eqref{eq:residue} above.

Next, setting the temporal components of $k_{\sf in}$ and $k_{\sf out}$ on shell, we can express the remaining dependence as $\mathcal{M}_a(\vec{k}_{\sf in}^2,\hat{\vec k}_{\sf in},\vec{k}_{\sf out}^2,\hat{\vec k}_{\sf out})$. Then the first set of angular momentum indices are defined by decomposing this amplitude in the usual way
\begin{align}
\label{eq:first_decom_M}
\mathcal{M}_a(\vec{k}_{\sf in}^2,\hat{\vec k}_{\sf in},\vec{k}_{\sf out}^2,\hat{\vec k}_{\sf out})= [\mathcal{M}_a]_{l'm',lm}(\vec{k}_{\sf in}^2,\vec{k}_{\sf out}^2)\ 4\pi |\vec{k}_{\sf in}|^l|\vec{k}_{\sf out}|^{l'} Y^*_{l'm'}(\vec{\hat{k}}_{\sf in})Y_{lm}(\vec{\hat{k}}_{\sf out}).
\end{align}

Next one makes the on-shell replacement $\boldsymbol k^2 \to p^2-a^2 \f(\boldsymbol k)$ which introduces additional angular dependence. This is then decomposed as in eq.~\eqref{eq:second_decom_LR} to reach
\begin{align}
\label{eq:second_decom_M}
[\mathcal{M}_a]_{l'm',lm}(\vec{k}^2_{\sf in},\vec{k}^2_{\sf out})& \to [\mathcal{M}_a]_{l'm',lm}(p^2-a^2 \f(\boldsymbol k_{\sf in}),p^2-a^2 \f(\boldsymbol k_{\sf out})) \,,\\[5pt]
&=[\mathcal{M}_a]_{l'm',lm',\alpha'\beta',\alpha\beta}(p^2,\vec{k}^2_{\sf in},\vec{k}^2_{\sf out})4\pi |\vec{k}_{\sf in}|^\alpha|\vec{k}_{\sf out}|^{\alpha'} Y^*_{\alpha'\beta'}(\hat{\vec{k}}_{\sf in})Y_{\alpha\beta}(\hat{\vec{k}}_{\sf out}) \,.
\nonumber
\end{align}

The on-shell momentum $p^2$ does not need an ``in'' or ``out'' label as it is determined by the finite-volume energy level $E$. The remaining $\vec{k}^2$ dependencies occur at order $a^2$ and hence can be placed on-shell via $\boldsymbol k^2\to p^2$. This can all be written as a single energy argument
\begin{align}
[\mathcal{M}_a]_{l'm',lm,\alpha'\beta',\alpha\beta}(E)=[\mathcal{M}_a]_{l'm',lm,\alpha'\beta',\alpha\beta}(p^2,p^2,p^2) \bigg \vert_{p^2 =E^2/4-m_\pi^2} \,.
\label{eq:final_M_rest}
\end{align}

This concludes the main result of this work. To summarize, we have defined a discretized version of the scattering amplitude with a transparent (albeit complicated) relation to the vertices of the SET. We have shown how this quantity can be extracted from the finite-volume spectrum for a calculation performed at a fixed lattice spacing. The expectation is that this quantity will have a more controlled continuum limit and that one might potentially make use of discretized chiral perturbation theory or other EFTs to fit the lattice artifacts in a more controlled way.

\section{Special cases of the general formalism}
\label{sec:simplified_results}

We now turn to special cases of the general formalism. First, we describe the truncation of the quantization to finite matrices, which follows from neglecting higher angular momentum contributions within $\mathcal M_a(E)$. We then turn to a perturbative expansion of our result for the case of weakly interacting low-energy degrees of freedom. This leads to simplified expressions for finite-volume energies with discretization effects and gives some insight into the extent to which lattice artifacts cancel between interacting and non-interacting energies. Finally we present perturbative results obtained in $\lambda \phi^4$ theory. This serves to illustrate the definition of the discretized scattering amplitude in our framework, and the approach for calculating the latter perturbatively, e.g.~in effective field theory.

\subsection{Truncation of the quantization condition}

As in the continuum case, the quantization condition can only be used in practice after truncation. In the present context, this amounts to selecting a pair of values $l_{\sf max}, \alpha_{\sf max}$ and formally setting $[\mathcal{M}_a]_{l'm',lm,\alpha'\beta',\alpha\beta}(E) = 0$ whenever $l,l'>l_{\sf max}$ or $\alpha,\alpha'>\alpha_{\sf max}$. For the case of $l,l'$ the truncation is motivated by the fact that the higher angular momentum contributions are suppressed by powers of the back-to-back momentum $p$, which translates to powers of $1/L$ in the case of weak interactions. For the $\alpha,\alpha'$ truncation, all non-zero components are suppressed by powers of $a^2$ and additional kinematic suppression arises as we describe in section~\ref{sec:perturbative_results}.

We first consider the restriction to $\alpha_{\sf max} = 0$, which will likely be sufficient to describe discretization effects in many lattice QCD calculations. This reduces the quantization condition to a single set of angular momentum indices
\begin{equation}
\label{eq:alpha_truncation}
\det_{lm} \left [\mathcal{M}_a(E)^{-1}-\boldsymbol{F}_{\!a}(E,L) \right ] = 0 \,,
\end{equation}
where $\boldsymbol{F}_{\!a}$ is given by
\begin{equation}
[\boldsymbol{F}_{\!a}]_{lm,l'm'}(E, L) = -\frac{1}{4 E} \left[\frac{1}{L^3}\sum_{\vec k}-\int \frac{d^3 \vec{k}}{(2\pi)^3}\right] \frac{4 \pi \vert \boldsymbol k \vert^{l+l'} Y_{lm}(\hat {\boldsymbol k}) Y_{l'm'}(\hat {\boldsymbol k}) \, {\sf H}(\vec{k})}{p^2-\vec{k}^2-a^2 \f (\vec{k}) + i \epsilon} \,.
\end{equation}
In this form, $\mathcal{M}_a(E)$ is simply given by the $
\alpha=\alpha'=\beta=\beta'=0$ component of the scattering amplitude in the construction detailed above, i.e.~one only keeps components that are non-zero in the continuum limit.

In the remainder of this section we will additionally consider the case of $l_{\sf max} = 0$. This allows us to drop the determinant and write the quantization condition as
\begin{equation}
\Ms (E)^{-1} - \Fs(E, L) = 0 \,,
\label{eq:S_wave_truncation}
\end{equation}
where we have introduced the shorthand
\begin{align}
\Ms (E) & = [\mathcal M_a]_{00,00,00,00}(E) \,,
\\
\Fs(E, L) & = [\boldsymbol F_a]_{00,00,00,00}(E, L) \,.
\end{align}
As we discuss in the following, the imaginary parts cancel in eq.~\eqref{eq:S_wave_truncation} so that one can also take the real part of the without any loss of information.

\subsection{Expansions of the low-lying finite-volume energies}

We now relate the finite-volume energies to scattering amplitude parameters for the case of weakly interacting low-energy degrees of freedom. We will find that the leading-order result, given in eq.~\eqref{eq:simp_lowest_E}, takes the same form as the corresponding continuum relation, provided one uses the appropriate single-particle energy everywhere.

Following the discussion in section~\ref{sec:simp_prop}, we first recall that the single-particle energies with total momentum $\boldsymbol k$ are given by
\begin{align}
\Omega_k=\sqrt{\vec{k}^2+m_\pi^2+a^2\f(\vec{k})} \,,
\end{align}
where $\vec{k}$ is quantized such that $\vec{k} \in (2\pi/L) \mathbb{Z}^3$.

We next introduce the label $N$ to denote a particular two-particle non-interacting energy, built from summing two $\Omega_k$ contributions, which must be equal since $\boldsymbol P = \boldsymbol 0$. Starting with the three-momenta, it will prove convenient to define $S_N$ as the set of all $\vec{k}$ that satisfy
\begin{align}
\label{eq:3_momentum_crit}
\vec{k}^2 = N (2\pi/L)^2 \,,
\end{align}
and define $\vec{k}_N$ as any representative solution in the set. We also denote the multiplicity of solutions (equivalently the number of elements of $S_N$) by $\eta_N$, e.g.~$\eta_0 = 1, \eta_1 = 6,$ $\eta_2 = 12, \cdots$.

For $N < 9$, this label is sufficient to indicate a particular single-particle energy:
\begin{align}
\Omega_N(L,a)=\sqrt{N (2 \pi/L)^2 + m_\pi^2+a^2\f(\vec{k}_N)} \,.
\end{align}
The key point is that, provided $N < 9$, all elements of $S_N$ are related by cubic rotations and reflections. This ensures that $\f(\vec{k})$ evaluates to the same result for every possible element, so that $\f(\boldsymbol k_N)$ is unambiguous. An ambiguity does arise at $N=9$ since both $\vec{k}=(2\pi/L)(0,0,3)$ and $\vec{k}=(2\pi/L)(2,2,1)$ solve eq.~\eqref{eq:3_momentum_crit}, but will generally yield different values of $\f(\boldsymbol k_N)$ and thus distinct energies. It is possible to accommodate this splitting, see e.g.~ref.~\cite{Grabowska:2021xkp}, but for the present discussion we restrict attention to $N<9$.\footnote{See also ref.~\cite{Peterken:2023zwu} for a discussion of such accidental degeneracies in the context of finite-volume matrix elements.}

To derive a result for two-particle energies, we next require a parameterization of the scattering amplitude. Here we take the standard choice of describing $S$-wave scattering with a single parameter
\begin{align}
\label{eq:Mparam_deff}
\mathcal{M}_a^{\sf S}(E)= \frac{16\pi E}{-a_0(a)^{-1}-i \big ( p + a^2 B_a(p) \big )} \,.
\end{align}
Here $a_0(a)$ is a generalization of the scattering length with lattice-spacing dependence. An example of the form that this might take is given in the context of $\lambda \phi^4$ theory in the following section.

We have also introduced $B_a(p)$, which can be determined from the imaginary part of $\boldsymbol F_a(E,L)$, following from the requirement that this should cancel the imaginary part of $\mathcal M_a(E)^{-1}$ in eq.~\eqref{eq:S_wave_truncation}. We are not concerned with the detailed form of this function since it does not contribute to the quantization condition and also vanishes in the continuum limit.

In the weakly interacting theory, the finite-volume energies are close to their non-interacting values and, as a result, when $\Fs(E, L)$ is evaluated at a solution, $p^2$ will be near a non-interacting momentum. This will lead to a dominate contribution to the sum that must be separated out while performing the expansion. More precisely, consider a particular energy $E_N(L, a)$ which, in the non-interacting limit, is given by
\begin{align}
E_N^{(0)}(L, a) = 2 \, \Omega_N (L, a) \,.
\end{align}
To describe the perturbative shifts to this energy, it is convenient to introduce $\delta p^2=p^2-\vec{k}_N^2+a^2\f(\vec{k}_N)$. Then $\Fs$ can be decomposed as
\begin{align}
\Fs(E, L) & =-\frac{\eta_N}{4 E L^3} \frac{1}{\delta p^2} + \deltaFs(E, L) \,,
\label{eq:f_lowest_order}
\end{align}
where the correction term, $\deltaFs(E, L)$, is given by
\begin{equation}
\deltaFs (E, L) =-\frac{1}{4E} \bigg[\frac{1}{L^3}\sum_{\vec k \not \in S_N}-\int \frac{d^3\vec k}{(2\pi)^3} \bigg] \frac{{\sf H}(\vec{k})}{p^2-\vec k^2-a^2\f(\vec{k})+i\epsilon} \,.
\end{equation}
Because all elements of $S_N$ are excluded from the sum defining $\deltaFs(E, L)$, this quantity is finite for $\delta p^2 \to 0$, and is therefore subdominant in a small $\delta p^2$ expansion.

\subsubsection{Leading order}

To complete the leading-order expansion we postulate, and then confirm {\em a posteriori} that $O(\delta p^2) = O(a_0)$. Our aim is therefore to determine the form of the quantization condition given in eq.~\eqref{eq:final_QC} to leading order in both of these parameters. We begin by expanding $\Ms$ in powers of $a_0$
\begin{align}
\mathcal{M}_a^{\sf S}(E)=-16\pi E a_0(a) + O(a_0^2) \,.
\end{align}

Substituting this expansion and eq.~\eqref{eq:f_lowest_order} into the quantization condition then gives
\begin{align}
0=-\frac{1}{16\pi E a_0(a)} + \frac{\eta_N}{4 E L^3}\frac{1}{\delta p^2} +O\big [ (a_0)^0 \big ] \,.
\end{align}
This is solved by $\delta p^2 = 4\pi \eta_N a_0(a)/ L^3$ and combining this result with the relation between $\delta p^2$ and $p^2$ we find that the solution in terms of the latter coordinate, which we denote by $p_N^2(L,a)$, takes the form
\begin{align}
p_N^2(L,a) = \vec{k}_N^2 + a^2 \f(\vec{k}_N) + \frac{4\pi \eta_N a_0(a)}{L^3} + O(a^3, a_0^2) \,.
\end{align}
The conversion to $E_N(L,a)^2$ is then achieved by adding $m_\pi^2$ and multiplying by 4 on both sides. Here one finds that the $\f(\boldsymbol k_N)$ dependence is then completely absorbed into the non-interacting energy:
\begin{align}
E_N(L,a)^2 = 4\Omega_N(L,a)^2 + \frac{16\pi \eta_N a_0(a)}{L^3} + O(a^3, a_0^2)\,.
\end{align}
Taking the square root of this expression, and expanding again to the same order, we find the leading-order result for the finite-volume energy levels:
\begin{align}
\label{eq:simp_lowest_E}
E_N(L,a) = 2\Omega_N(L,a) + \eta_N \frac{4\pi a_0(a)}{\Omega_N(L,a) L^3} +O(a^3,a_0^2) \,.
\end{align}
This matches the form of the continuum expansion as given in refs.~\cite{\leadingEnergyExpansion} when $a \to 0$.

This result has some significance for the analysis of finite-volume energies in lattice QCD computations. In particular, the difference $E_N(L,a) - 2 \Omega_N(L,a)$ is extracted when one forms a ratio of variational eigenvalues to products of single-hadron correlators. Thus, this expansion shows that, for weakly interacting systems, this ratio cancels certain discretization effects.

\subsubsection{Higher orders}

The extension to higher orders is also possible, and closely follows the approach of refs.~\cite{Luscher:1986pf,Beane:2007qr,Hansen:2016fzj,Grabowska:2021xkp}. First we make use of the fact that the imaginary part of $\Ms(E)^{-1}$ must cancel with the imaginary part of $\Fs(E, L)$ for the quantization condition to have continuous solutions. We can therefore restrict attention to the real parts given by
\begin{align}
\text{Re} \left [ \mathcal{M}_a^{\sf S}(E)^{-1} \right ] & =-\frac{1}{16\pi E a_0(a)} \,,
\\
\text{Re} \left [ \Fs(E, L) \right ] & =-\frac{\eta_N}{4 E L^3} \frac{1}{\delta p^2} + \deltaFsPV(E, L) \,,
\end{align}
where we have introduced
\begin{align}
\deltaFsPV (E, L) & = \frac{1}{4E} \bigg[\frac{1}{L^3}\sum_{\vec k \not \in S_N}-{\sf pv} \int \frac{d^3\vec k}{(2\pi)^3} \bigg] \frac{{\sf H}(\vec{k})}{\vec k^2 + a^2\f(\vec{k})-p^2} \,.
\end{align}
This contribution, equal to the real part of $\deltaFs$, was irrelevant at leading order but will contribute at higher orders, leading to the first nontrivial $a$ dependence in the expansion.

To see this, we expand in powers of $\delta p^2$ to write
\begin{align}
\deltaFsPV (E, L) = \frac{1}{16 \pi E} \sum_{n=0}^\infty (\delta p^2)^n C_n(L, a^2, N) \,,
\end{align}
where
\begin{align}
\hspace{-6pt} C_n(L, a^2, N) & = \frac{4 \pi}{n!} \lim_{x \to {\vec k}_N^2 + a^2 f(\boldsymbol k_N)} \frac{\partial^n}{\partial x^n} \bigg[\frac{1}{L^3}\sum_{\vec k \not \in S_N}-{\sf pv} \int \frac{d^3\vec k}{(2\pi)^3} \bigg] \frac{{\sf H}(\vec{k})}{\vec k^2 + a^2\f(\vec{k})-x} \,, \\[5pt]
& \hspace{0pt} = 4 \pi \bigg[\frac{1}{L^3}\sum_{\vec k \not \in S_N}-{\rm Re} \int \frac{d^3\vec k}{(2\pi)^3} \bigg] \frac{{\sf H}(\vec{k})}{\left [ \vec k^2-{\vec k}_N^2 + a^2[\f(\vec{k})-\f(\vec{k}_N)] - i \epsilon \right ]^{n+1}} \,.
\end{align}
In the last step we have re-expressed the pole prescription in terms of the real part of the $i \epsilon$ integral. This is required to commute the differentiation with the integral.

Substituting this higher-order expansion into eq.~\eqref{eq:final_QC}, one reaches a form that can be solved order by order
\begin{equation}
-\frac{1}{a_0(a)} + \frac{4 \pi \eta_N}{L^3} \frac{1}{\delta p^2}-\sum_{n=0}^\infty (\delta p^2)^n C_n(L, a^2, N) = 0 \,.
\end{equation}
For example, the next-to-leading order expression is given by
\begin{equation}
p_N^2(L,a) = \vec{k}_N^2 + a^2 \f(\vec{k}_N) + \eta_N \frac{4\pi a_0(a)}{ L^3} \left [1 - \frac{a_0(a)}{\pi L} \mathcal I(L, a^2, N) \right ] + O(a_0^3, a^3) \,,
\end{equation}
where $\mathcal I(L, a^2, N)$ is a generalization of the standard coefficient of refs.~\cite{Huang:1957im,Beane:2007qr}
\begin{equation}
\mathcal I(L, a^2, N)= \bigg[ \sum_{\vec n \not \in \widetilde S_N}-{\sf pv} \int \! d^3\vec n \, \bigg] \frac{{\sf H}(2 \pi \vec{n}/L)}{ \vec n^2-{\vec n}_N^2 + a^2 L^2 [\f(2 \pi \vec{n}/L)-\f(2 \pi \vec{n}_N/L)]/(2 \pi)^2 } \,,
\end{equation}
with $\widetilde S_N = [L/(2 \pi)] S_N$.

In the continuum limit ($a \to 0$) and for the ground state ($N=0$) case, this quantity has only exponentially suppressed $L$-dependence entering through $\sf H$. In this case, one recovers the well known geometric constant~\cite{Huang:1957im,Beane:2007qr}
\begin{equation}
\mathcal I = \lim_{L \to \infty} \mathcal I(L, 0,0) = \lim_{\Lambda \to \infty} \left [ \sum_{\boldsymbol n \neq \boldsymbol 0}^{ \vert \boldsymbol n \vert < \Lambda} \frac{1}{\boldsymbol n^2} - 4 \pi \Lambda \right ] = - 8.914 \cdots \,.
\end{equation}
See also the discussion in appendix A of ref.~\cite{Hansen:2016fzj}.

\subsection{Explicit results for \texorpdfstring{$\lambda\phi^4$}{phi-four} theory}
\label{sec:perturbative_results}

In this section we present explicit results in the case of a scalar $\phi^4$ theory. We start by calculating the single-particle propagator and scattering amplitude in perturbation theory and then calculate the finite-volume energy levels and show their dependence on the lattice spacing.

The relevant Lagrangian, including discretization effects of $O(a^2)$, is as follows \cite{Symanzik:1983dc}:
\begin{align}
\mathcal L = \frac 12 (\partial_\mu \phi)^2 + \frac 12 m_{\sf c}^2 \phi^2 +\frac {\lambda}{4!} \phi^4-\frac{1}{2} \frac{a^2}{12} \phi [\partial]^4\phi \,,
\end{align}
where we only keep vertices that contribute to two an four-point functions at leading order.
In the following, we will only keep $S$-wave effects and expand to leading order in $\lambda$ and to next-to-leading order, $O(a^2)$, in the lattice spacing.

The leading order single-particle propagator is given by:
\begin{align}
D_a(k)=\frac{1}{k^2-\frac{1}{12} a^2 [k]^4 + m_{\sf c}^2} \,.
\end{align}
Following the definitions of section~\ref{sec:simp_prop}, we first note
\begin{align}
f(k) =-\frac{1}{12} \left ( [\vec{k}]^4 + k_4^4 \right ) \,, \qquad \qquad % \\
z_a(k) = 1 \,.
\end{align}
From here it follows that $f(i \omega_k, \vec{k}) =-\frac{1}{12} \left ( [\vec{k}]^4 + \omega_k^4 \right )$ and thus
\begin{equation}
{\sf f}(\boldsymbol k) = f(i\omega_k,\vec{k})-f(i\omega_0,\vec{0}) =-\frac{1}{12} \left ( [\vec{k}]^4 + (\boldsymbol k^2)^2 + 2 \boldsymbol k^2 m_{\pi}^2 \right ) \,,
\end{equation}
where we recall that $[\vec{k}]^4=k_x^4+k_y^4+k_z^4$, and note that this introduces angular dependence. We additionally have
\begin{align}
Z_a(k)
= \left [ 1-\frac{a^2}{12} \frac{k_4^4-\omega_k^4}{k_4^2 + \omega_k^2} \right ]^{-1} \qquad \Longrightarrow \qquad
\Z (\vec{k}) = 1-\frac{a^2}{6} \big ( \boldsymbol k^2 +m_{\pi}^2 \big )
\,.
\end{align}
All the quantities are written in terms of $m_\pi$, which we recall is the zero-momentum single-hadron energy at a given value of the lattice spacing.

At leading order in perturbation theory, the scattering amplitude is given by
\begin{align}
[\mathcal{M}_a]_{00,00}(\vec{k}_{\sf in}^2,\vec{k}_{\sf out}^2) =-\lambda {\sf Z}_a(\vec{k}_{\sf in}){\sf Z}_a(\vec{k}_{\sf out}) =-\lambda+a^2\lambda\frac{1}{6}(\vec{k}_{\sf in}^2+\vec{k}_{\sf out}^2+2m^2_{\pi}) \,,
\end{align}
where the second term arises from the $a$-dependent wave-function renormalization. Here we have already used the fact that the result has no directional dependence to identify this as the $00,00$ component of the scattering amplitude defined in eq.~\eqref{eq:first_decom_M}.

Following the procedure of eq.~\eqref{eq:second_decom_M}, the next step is to make the replacements
\begin{equation}
\vec{k}_{\sf in}^2 \to p^2-a^2 \f(\vec k_{\sf in}) \qquad \text{and} \qquad \vec{k}_{\sf out}^2 \to p^2-a^2 \f(\vec k_{\sf out}) \,.
\end{equation}
However, since the momentum dependence only enters at $O(a^2)$, the additional directional dependence arising within $\f(\boldsymbol k)$ is beyond the order we control. We thus find the fully isotropic expression
\begin{equation}
\mathcal{M}_a^{\sf S}(p^2) =-\lambda+a^2\lambda\frac{1}{3}(p^2+m^2_{\pi})\,,
\end{equation}
where we recall $\mathcal{M}_a^{\sf S}(p^2) \equiv [\mathcal{M}_a]_{00,00,00,00}(p^2)$ and that $O(a^3)$ effects are dropped throughout.

We further find it convenient to express the result in terms of the scattering length $a_0(a)$, defined as
\begin{equation}
\mathcal{M}_a^{\sf S}(p^2) \bigg \vert_{p^2 = 0} = - \lambda \left [ 1 - \frac13 (m_\pi a)^2 \right ] =-32\pi m_\pi a_0(a) \,.
\end{equation}
Using this to relate the coupling to $a_0(a)$ and then expressing the scattering amplitude in terms of the latter, we deduce
\begin{align}
\mathcal{M}_a^{\sf S}(p^2) =-32\pi m_\pi a_0(a) \left [1-\frac{a^2 p^2}{3} \right ] \,.
\end{align}
Observe that, at leading order in perturbation theory, the scattering amplitude is real and hence the imaginary part of eq~\eqref{eq:Mparam_deff} is effectively set to zero.

This can now be used to predict the finite-volume spectrum for any values of $a_0(a)$ and $L$. This requires the numerical evaluation of the $S$-wave element of $\boldsymbol{F}_a$
\begin{align}
\Fs(E, L) =-\frac{1}{2}\left[\frac{1}{L^3}\sum _k-\int \frac{d^3 \vec{k}}{(2\pi)^3}\right] \frac{1}{2E}\frac{{\sf H}(\vec{k})}{p^2-\vec{k}^2-a^2\f(\vec{k})+i\epsilon} \,.
\end{align}
Here it is convenient to decompose $\f(\boldsymbol k)$ in spherical harmonics as
\begin{equation}
\f(\boldsymbol k) = \sum_{\ell \in \{0, 4\}} \sum_{m=-\ell}^{\ell} \f_{\ell m}(k) \sqrt{4 \pi} Y_{\ell m}(\hat{\boldsymbol k}) \,,
\end{equation}
where all $\f_{\ell m}(k)$ vanish besides the following set:
\begin{align}
\f_{00}(k) & =-\frac{1}{12} \left ( \frac{8}{5} k^4 + 2 k^2 m_{\pi}^2 \right ) \,, \\
\f_{40}(k) & =-\frac{1}{90} k^4 \,, \\
\f_{4\pm 4}(k) & =-\frac{1}{90} \sqrt\frac{5}{14} k^4 \,.
\end{align}

In the evaluation presented here we further simplify the procedure by only keeping the $S$-wave contribution to $\f(\boldsymbol k)$.
This approximation is distinct from the truncations of $\mathcal M_a(E)$, but is again motivated by barrier factor suppression. In particular we note that, in the $k\to0$ limit, the $\f_{00}(k)$ dominates as the ratios scale as
\begin{align}
\frac{\f_{40}(k)}{\f_{00}(k)} = O(k^2) \,, \qquad \frac{\f_{4\pm4}(k)}{\f_{00}(k)} = O(k^2)\,.
\end{align}
In addition, in the $k\to\infty$ limit, the ratios tend to fixed values less than one
\begin{align}
\lim_{k \to \infty} \frac{\f_{40}(k)}{\f_{00}(k)} = \frac{1}{12} \,, \qquad
\lim_{k \to \infty} \frac{\f_{4\pm4}(k)}{\f_{00}(k)} = \frac{\sqrt{70}}{168} \,,
\end{align}
so that even in the momentum region where the overall contribution is suppressed, the $G$-wave components of $\f(\boldsymbol k)$ are subdominant.

As an aside, we briefly comment that dropping the higher-partial waves of $\f(\boldsymbol k)$ also allows one to discard the unphysical pole in $\boldsymbol k$ by factorizing the isotropic pole in $k^2$ into two terms and evaluating the unphysical factor at the position of the physical pole. This has the advantage that one can extend the cutoff function to arbitrarily high momenta. We do not pursue this additional reduction for our evaluations here, as we find the results with the smooth cutoff to be numerically stable and sufficient for the present illustrative purposes.

Applying the $S$-wave truncation of $\f(\boldsymbol k)$ to $\Fs$, we find
\begin{align}
\Fs(E, L) & \to -\frac{1}{4 E}\left[\frac{1}{L^3}\sum _k-\int \frac{d^3 \vec{k}}{(2\pi)^3}\right] \frac{{\sf H}(\vec{k})}{p^2-\vert \vec{k} \vert^2+a^2 \left [ \frac{2}{15} \vert \vec{k} \vert^4+\frac{1}{6}m^2 \vert \vec{k}\vert^2 \right ]+i\epsilon} \,, \\
& \hspace{-10pt} = -\frac{1}{4 E L (2\pi)^2}\left[\sum _{\boldsymbol n \in \mathbb{Z}^3}-\int d^3 \boldsymbol n\right] \frac{{\sf H}(2 \pi \vec{n}/L)}{q^2- \vert \boldsymbol n \vert^2+ (m_\pi a)^2 \left[\frac{8\pi^2 \vert \vec{n} \vert^4}{15 (m L)^2}+\frac{\vert\vec{n}\vert^2}{6}\right]+i\epsilon} \,,
\end{align}
where in the second line we have defined $\boldsymbol k=\frac{2\pi}{L} \boldsymbol n$ and $p=\frac{2\pi}{L} q$.

\begin{figure}
\centering
\includegraphics[scale=0.65]{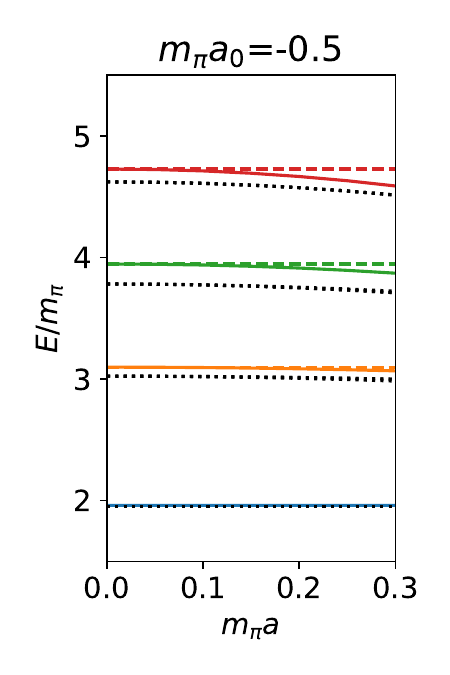}
\hspace{-15pt}
\includegraphics[scale=0.65]{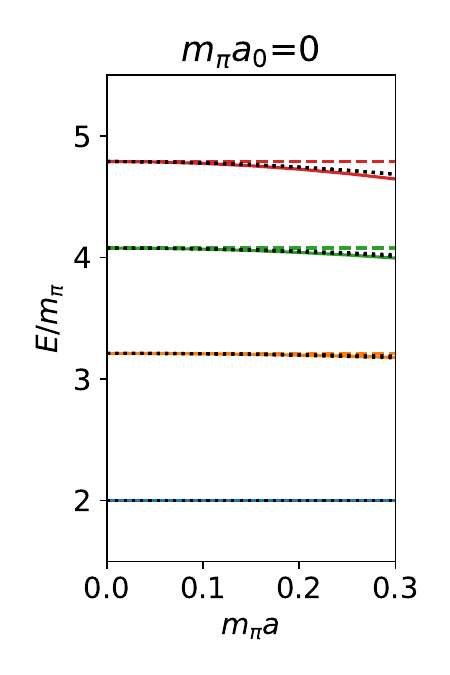}
\hspace{-15pt}
\includegraphics[scale=0.65]{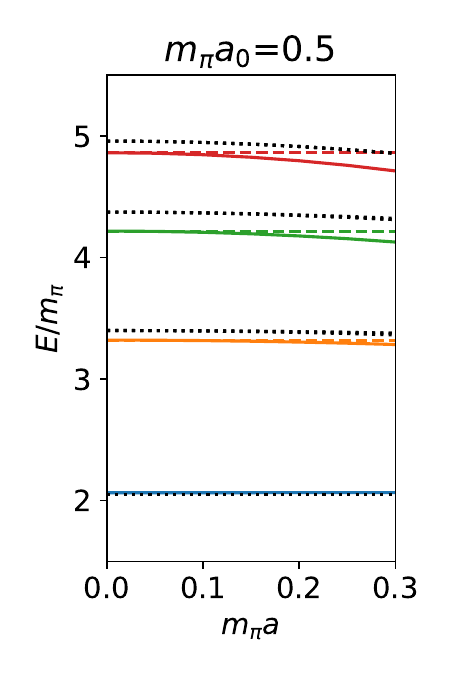}
\caption{Finite-volume energies with discretization effects vs.~the lattice spacing $m_\pi a$ at fixed physical volume, $m_\pi L=5$. The solid colorful lines show the $a$-dependent energies while the dashed colorful lines show the same in the continuum. We also include the discretized non-interacting energies as dotted black curves for reference.}
\label{fig:disc_eff_plot}
\end{figure}

Finally making use of the fact that the imaginary parts of eq.~\eqref{eq:S_wave_truncation} cancel in general, and that the imaginary part of $\Ms$ vanishes to the order we work, we write the quantization condition as
\begin{align}
\left(-32\pi m_\pi a_0(a) \left [1-\frac{a^2 p^2}{3} \right ]\right)^{-1} = \text{Re} \, \Fs(E, L) \,.
\end{align}
In figure~\ref{fig:disc_eff_plot} we show results from solving this numerically, resulting in predictions for the dependence of the finite-volume energy levels ($E_n(L,a)/m_\pi$) on the lattice spacing ($m_\pi a$), for fixed values of the scattering length and box size ($m_\pi a_0$ and $m_\pi L$). These plots also include the non-interacting energies (with discretization effects) for reference.

\section{Conclusion}
\label{sec:conclusion}

In this work we have extended the quantization condition of refs.~\cite{Luscher:1986pf,Rummukainen:1995vs,Kim:2005gf}, which relates finite-volume energies to $2\to2$ scattering amplitudes, to include the effects of non-zero lattice spacing $a$. As summarized in eqs.~\eqref{eq:f_disc_final} and \eqref{eq:final_QC}, the key result is a new quantization condition that depends on the lattice spacing in two ways: First, a dependence emerges in our extension of the finite-volume function (eq.~\eqref{eq:f_disc_final}) which incorporates the discretization effects of the single-hadron states. This leads to a new set of angular momentum indices in the quantization condition, encoding the $a$-induced angular momentum mixing. Second, the scattering amplitude itself is modified by the lattice spacing, but with a theoretically clear definition given in eq.~\eqref{eq:adep_scattering_amp}.

The utility of our new formalism may not be immediately obvious, since it yields a lattice-spacing-dependent scattering amplitude, a feature that is also technically true of the original quantization conditions. One key difference is that the definition of our fixed-$a$ scattering amplitude is transparent in the context of Symanzik Effective Theory, potentially making parameterization and extrapolation of numerical data tractable. A second key feature, already mentioned in the introduction, is that our approach gives a consistent version of the scattering amplitude over all values of the total momentum on a given lattice ensemble. As a result, one can reliably extract the lattice-spacing dependent scattering parameters, ensemble by ensemble, before considering the continuum limit.

To better understand our main result, in section~\ref{sec:simplified_results} we have studied several checks and limiting cases. First, we considered the weakly interacting expansion of the lowest two-particle energy level in terms of a lattice-spacing dependent scattering length. Second, we applied our formalism to $\lambda\phi^4$ theory to demonstrate how the $a$-dependent scattering amplitude can be computed in practice.

We emphasize that our formalism is only relevant if the discretization effects can be resolved for single-hadron states (or otherwise theoretically parameterized, e.g. using effective field theory). For this reason, the formulas may be most useful in heavy-quark studies, where the single-hadron discretization effects are easier to resolve. Concrete examples in this direction are $D \pi$ or $DD$ scattering, both of which have already been investigated in various lattice QCD calculations~\cite{\heavyquark}.

A simple limiting case of our method arises in the situation where, for a given continuum trajectory, lattice artifacts can be resolved in the particle mass (i.e.~in the $a$-dependence of the single-hadron rest-frame energy), but not in the dispersion relation at fixed $a$. In this case, our method reduces to the intuitive approach of using the fixed-$a$ mass in the quantization condition, as is commonly done in lattice QCD studies.

We conclude with two possible avenues of future work for this formalism:

The first is the treatment of more complicated discretization effects. For example, the work of refs.~\cite{Husung:2021mfl,Husung:2022kvi} has cataloged logarithmic discretization effects arising from the running of the coupling. Another interesting extension is that to both rooted and unrooted staggered quarks. In these cases, the single-hadron dispersion relation is modified by the presence of taste-breaking terms, and it will be necessary to incorporate the multiple tastes into the formalism such that one will recover a coupled-channel approach that reduces to the physical scattering channels for $a \to 0$. In the rooted case, other challenges may arise from the nonunitary nature of the theory. We will not speculate here on how prohibitive these challenges might be.

A second direction of future work is to mimic the developments of the continuum finite-volume formalism, for example the extension to twisted boundary conditions, to multiple two-particle channels with non-identical and non-degenerate particles, and to intrinsic spin. The latter is particularly interesting in light of the results of ref.~\cite{Green:2021qol}. The main complication in the context of spin will be to write the momentum-space Dirac operator at fixed $a$ in terms of spin states, and to quantitatively encode the discretization effects of the spinors in the definition of $\boldsymbol F_a$. Other natural extensions include working out the formalism for $1 \overset{\mathcal J}{\to} 2$ and $2 \overset{\mathcal J}{\to} 2$ transition amplitudes, as well as for three-particle scattering.

\acknowledgments

We thank Jeremy Green and Fernando Romero-L\'opez for useful discussions. Both TP and MTH and supported in part by UK STFC grants ST/P000630/1 and ST/X000494/1. MTH is additionally supported by UKRI Future Leader Fellowship MR/T019956/1.

\appendix
\section{Non-zero momentum}
\label{app:non_zero_mom}

In this appendix we generalize eq.~\eqref{eq:final_QC} to accommodate non-zero spatial momentum in the finite-volume frame. Our starting point is the appropriate modification of eq.~\eqref{eq:LFR_disc}
\begin{align}
L_a \boldsymbol{F}_a R_a &= \frac{1}{2}\int \frac{dk_4}{2\pi}\left[\frac{1}{L^3}\sum_{\vec k}-\int \frac{d^3 \vec{k}}{(2\pi)^3}\right] \frac{L_a(P, k) \, {\sf H}(\boldsymbol P, \boldsymbol k) \, R_a(P,k)}{[k_4^2 + \Omega_k^2] [(P_4-k_4)^2 + \Omega_{P-k}^2]} \,,
\end{align}
where we recall and introduce
\begin{align}
\Omega_{k} = \sqrt{\boldsymbol k^2 + m_\pi^2 + a^2 {\sf f}(\boldsymbol k)} \,, \qquad
\Omega_{P-k} & = \sqrt{(\boldsymbol P-\boldsymbol k)^2 + m_\pi^2 + a^2 {\sf f}(\boldsymbol P-\boldsymbol k)} \,.
\end{align}
The function ${\sf H}(\boldsymbol P, \boldsymbol k)$ is an extension of ${\sf H}(\boldsymbol k)$, discussed in detail below.

As in section~\ref{sec:disc_sum_int_diff}, the first step is to evaluate the $k_4$ integral and set $P_4 = i E$ to reach
\begin{align}
L_a \boldsymbol{F}_a R_a &=-\frac{1}{2} \left[\frac{1}{L^3}\sum_{\vec k}-\int \frac{d^3 \vec{k}}{(2\pi)^3}\right] \frac{{\sf L}_a(\boldsymbol P, \boldsymbol k) \ {\sf H}(\boldsymbol P, \boldsymbol k) \ {\sf R}_a(\boldsymbol P, \boldsymbol k)}{2 \Omega_k \left [(E-\Omega_k)^2-\Omega^2_{P-k} + i \epsilon \right ]} \,.
\end{align}
In the usual moving frame derivation of refs.~\cite{Rummukainen:1995vs,Kim:2005gf}, the next step is to identify the pole as a Lorentz scalar and to rewrite it using CMF coordinates. This step does not follow in the present approach, however, since $\f(\boldsymbol k)$ and therefore $\Omega_k$ include Lorentz-symmetry-violating terms. To make progress, we instead expand the pole to $O(a^2)$ and write
\begin{align}
L_a \boldsymbol{F}_a R_a &=-\frac{1}{2}\left[\frac{1}{L^3}\sum_{\vec k}-\int \frac{d^3 \vec{k}}{(2\pi)^3}\right]\frac{{\sf L}_a(\boldsymbol P, \boldsymbol k) \ {\sf H}(\boldsymbol P, \boldsymbol k) \ {\sf R}_a(\boldsymbol P, \boldsymbol k)}{2\Omega_k \left [ (E-\omega_k)^2-\omega_{P-k}^2-a^2 \f^{ (i)}(\boldsymbol P, \boldsymbol k) + i \epsilon \right ]} \,,
\end{align}
where we have introduced the intermediate function
\begin{equation}
\f^{(i)}(\boldsymbol P, \boldsymbol k) = \frac{(E-\omega_k)}{\omega_k} \f(\boldsymbol k) + \f(\boldsymbol P-\boldsymbol k) \,.
\end{equation}

At this stage it is possible to express quantities in terms of an alternative coordinate: $\boldsymbol k^*$. As in ref.~\cite{Kim:2005gf}, this is defined as the spatial part of $k^{* \mu}$, which is the result of boosting the Minkowski four-momentum $k^\mu = (\omega_k, \boldsymbol k)$ with velocity $\boldsymbol \beta =-\boldsymbol P/E$. Making use of this coordinate we first write
\begin{align}
L_a \boldsymbol{F}_a R_a &=-\frac{1}{2}\left[\frac{1}{L^3}\sum_{\vec k} \frac{\omega_k^*}{\omega_k}-\int \frac{d^3 \vec{k}^*}{(2\pi)^3} \right] \frac{\omega_k}{\omega_k^*} \frac{{\sf L}_a(\boldsymbol P, \boldsymbol k) \ {\sf H}(\boldsymbol k^{*}) \ {\sf R}_a(\boldsymbol P, \boldsymbol k)}{2\Omega_k [(E^*-\omega_k^*)^2-\omega_{k}^{*2}-a^2 \f^{(i)}(\boldsymbol P, \boldsymbol k) + i \epsilon ]} \,,
\end{align}
where we have introduced
\begin{equation}
E^* = \sqrt{E^2-\boldsymbol P^2} \,, \qquad \qquad \omega_k^* = \sqrt{\boldsymbol k^{*2} + m_\pi^2} \,,
\end{equation}
and have expressed the cutoff function ${\sf H}$ in terms of $\boldsymbol k^*$. In terms of this coordinate, one can exactly use the definition given in eqs.~\eqref{eq:xdef} and \eqref{eq:Jdef} of the main text.

We further simplify by expanding the denominator, inserting $1 = (E^* + 2 \omega_k^*)/(E^* + 2 \omega_k^*)$, and setting $E^{*2}/4-\omega_k^{*2} = p^{*2}-\boldsymbol k^{*2}$ where
\begin{equation}
p^{*2} = E^{*2}/4-m_\pi^2 \,.
\end{equation}
This yields
\begin{align}
L_a \boldsymbol{F}_a R_a &=-\frac{1}{4 E^*}\left[\frac{1}{L^3}\sum_{\vec k} \frac{\omega_k^*}{\omega_k}-\int \frac{d^3 \vec{k}^*}{(2\pi)^3} \right] \frac{{\sf L}_a(\boldsymbol P, \boldsymbol k) \ {\sf H}(\boldsymbol k^{*}) \ {\sf R}_a(\boldsymbol P, \boldsymbol k)}{\mathcal C(\boldsymbol P, \boldsymbol k) \left [p^{*2}-\boldsymbol k^{*2}-a^2 \f^{(ii)}(\boldsymbol P, \boldsymbol k) + i \epsilon \right ]} \,,
\end{align}
where
\begin{align}
\f^{(ii)}(\boldsymbol P, \boldsymbol k) & = \left [ (E-\omega_k) \f(\boldsymbol k) + \omega_k \f(\boldsymbol P-\boldsymbol k) \right ] \frac{E^* + 2 \omega_k^*}{4 E^* \omega_k} \,, \\
\mathcal C(\boldsymbol P, \boldsymbol k) & = \frac{\omega_k^* 4\Omega_k}{\omega_k(E^* + 2\omega_k^*)} \,.
\end{align}

We continue to simplify, first by expanding $\Omega_k$ within $\mathcal C(\boldsymbol P, \boldsymbol k)$ to order $a^2$ to write
\begin{align}
\mathcal C(\boldsymbol P, \boldsymbol k) & = \frac{4 \omega_k^*}{E^* + 2\omega_k^*} \left( 1 + \frac{1}{2\omega_k^2} a^2 \f(\boldsymbol k) \right) \,,
\end{align}
and then by using the pole condition $p^{*2} = \boldsymbol k^{*2} + a^2 \f^{(ii)}(\boldsymbol P, \boldsymbol k)$ to expose all $a^2$ factors at fixed $E^*$. This gives
\begin{align}
\mathcal C(\boldsymbol P, \boldsymbol k) & = \left (1-\frac{a^2}{E^{*2}} \f^{(ii)}(\boldsymbol P, \boldsymbol k) \right ) \left( 1 + \frac{1}{2\omega_k^2} a^2 \f(\boldsymbol k) \right) \,,
\\
& = 1-a^2 \f(\boldsymbol k) \left [ (E-\omega_k) \frac{E^* + 2 \omega_k^*}{4 E^{*3} \omega_k}-\frac{1}{2 \omega_k^2} \right ]-a^2 \f(\boldsymbol P-\boldsymbol k) \frac{E^* + 2 \omega_k^*}{4 E^{*3}} \,,
\end{align}

Next, anticipating the final result, we define
\begin{equation}
\f_P(\boldsymbol k^*) = \frac12 \left [ \frac{E - \omega_k}{E - \omega_{P-k}} \f(\boldsymbol k) + \f(\boldsymbol P-\boldsymbol k) \right ] \,,
\end{equation}
and observe that
\begin{equation}
\mathcal C(\boldsymbol P, \boldsymbol k) [p^{*2}-\boldsymbol k^{*2}-a^2 \f^{(ii)}(\boldsymbol P, \boldsymbol k)] = \big (1 + \mathcal Q(\boldsymbol P, \boldsymbol k) \big ) [p^{*2}-\boldsymbol k^{*2}-a^2 \f_P(\boldsymbol k^*)] + O(a^4) \,,
\label{eq:remove_C}
\end{equation}
where we have further introduced
\begin{equation}
\mathcal Q(\boldsymbol P, \boldsymbol k) = \mathcal C(\boldsymbol P, \boldsymbol k) \left [ 1 - a^2 \frac{\f^{(ii)}(\boldsymbol P, \boldsymbol k) - \f_P(\boldsymbol k^*)}{p^{*2}-\boldsymbol k^{*2}} \right ] -1 \,.
\end{equation}
To complete the construction, we finally show that $\mathcal Q(\boldsymbol P, \boldsymbol k) = O(a^4)$, so that $\boldsymbol F_a$ can be simplified using eq.~\eqref{eq:remove_C}, with $\mathcal Q(\boldsymbol P, \boldsymbol k)$ set to zero to the order we work.

This follows from multiplying out and substituting to first write
\begin{equation}
\mathcal Q(\boldsymbol P, \boldsymbol k) = -a^2 \left [\f(\boldsymbol k) \, \mathcal Q_{\boldsymbol k}(\boldsymbol P, \boldsymbol k) + \f(\boldsymbol P-\boldsymbol k) \, \mathcal Q_{\boldsymbol P - \boldsymbol k}(\boldsymbol P, \boldsymbol k) \right ] \,,
\end{equation}
with
\begin{align}
\mathcal Q_{\boldsymbol k}(\boldsymbol P, \boldsymbol k) & = (E-\omega_k) \frac{E^* + 2 \omega_k^*}{4 E^{*3} \omega_k}-\frac{1}{2 \omega_k^2} + \frac{ (E-\omega_k) \frac{E^* + 2 \omega_k^*}{4 E^* \omega_k } - \frac12 \frac{E - \omega_k}{E - \omega_{P-k}} }{ p^{*2}-\boldsymbol k^{*2} } \,,
\\
\mathcal Q_{\boldsymbol P - \boldsymbol k}(\boldsymbol P, \boldsymbol k) & = \frac{E^* + 2 \omega_k^*}{4 E^{*3}} + \frac{ \frac{E^* + 2 \omega_k^*}{4 E^* } - \frac12 }{p^{*2}-\boldsymbol k^{*2}} \,.
\end{align}
It then only remains to show that $\mathcal Q_{\boldsymbol k}(\boldsymbol P, \boldsymbol k)$ and $\mathcal Q_{\boldsymbol P - \boldsymbol k}(\boldsymbol P, \boldsymbol k)$ are both $O(a^2)$ when evaluated at the pole.

We begin with $\mathcal Q_{\boldsymbol P - \boldsymbol k}(\boldsymbol P, \boldsymbol k)$, which turns out to be easier to reduce:
\begin{align}
\mathcal Q_{\boldsymbol P - \boldsymbol k}(\boldsymbol P, \boldsymbol k) & = \frac{E^* + 2 \omega_k^*}{4 E^{*3}} + \frac{ \frac{E^* + 2 \omega_k^*}{4 E^* } - \frac12 }{p^{*2}-\boldsymbol k^{*2}}
\\ & \to \frac{1}{2 E^{*2}} - \frac{ E^* - 2 \omega_k^*}{4 E^*(p^{*2}-\boldsymbol k^{*2})} + O(a^2)\,,
\\
& = \frac{1}{2 E^{*2}} - \frac{ E^{*2} - 4 \omega_k^{*2}}{4 E^* (E^* + 2 \omega_k^*)(p^{*2}-\boldsymbol k^{*2})} + O(a^2) \,,
\\
& = \frac{1}{2 E^{*2}} - \frac{ 1}{ E^* (E^* + 2 \omega_k^*)} + O(a^2)
\\ & \to O(a^2) \,,
\end{align}
where $\to$ indicates that the expression was simplified by evaluating at the pole.

Similarly for $\mathcal Q_{\boldsymbol k}(\boldsymbol P, \boldsymbol k)$, we begin by writing
\begin{align}
\mathcal Q_{\boldsymbol k}(\boldsymbol P, \boldsymbol k) & = (E-\omega_k) \frac{E^* + 2 \omega_k^*}{4 E^{*3} \omega_k}-\frac{1}{2 \omega_k^2} + \frac{ (E-\omega_k) \frac{E^* + 2 \omega_k^*}{4 E^* \omega_k} - \frac12 \frac{E - \omega_k }{E - \omega_{P-k}}}{(p^{*2}-\boldsymbol k^{*2})} \,,
\\
\begin{split}
& \to
(E-\omega_k) \frac{1}{2 E^{*2} \omega_k}-\frac{1}{2 \omega_k^2}
\\ & \hspace{60pt} + \frac{ (E-\omega_k) [(E - \omega_{P-k}) (E^* + 2 \omega_k^*) - 2 E^* \omega_k ]}{4 E^* \omega_k (E - \omega_{P-k})(p^{*2}-\boldsymbol k^{*2})} + O(a^2) \,.
\end{split}
\end{align}
The next step is to split the numerator into two terms, each of which vanish at the pole. Suppressing the $O(a^2)$, this can be written as
\begin{align}
\begin{split}
\hspace{-10pt} \mathcal Q_{\boldsymbol k}(\boldsymbol P, \boldsymbol k)
%%%
& \to
(E-\omega_k) \frac{1}{2 E^{*2} \omega_k}-\frac{1}{2 \omega_k^2}
\\ & \hspace{12pt} + \frac{(E-\omega_k)}{4 E^* \omega_k^2} \left [\frac{ (E - \omega_{P-k} - \omega_k) (E^* + 2 \omega_k^*) + \omega_k(E^* + 2 \omega_k^*) - 2 E^* \omega_k }{ (p^{*2}-\boldsymbol k^{*2})} \right ] \,,
\end{split}
\\[5pt]
\begin{split}
%%%
& \to
(E-\omega_k) \frac{1}{2 E^{*2} \omega_k}-\frac{1}{2 \omega_k^2}
\\ & \hspace{85pt} + \frac{(E-\omega_k)}{4 E^* \omega_k^2} \left [ \frac{ (E - \omega_{P-k} - \omega_k) 2E^* }{(p^{*2}-\boldsymbol k^{*2})} - \frac{\omega_k (E^* - 2 \omega_k^*) }{(p^{*2}-\boldsymbol k^{*2})} \right ] \,,
\end{split}
\\[5pt]
\begin{split}
%%%
& \to
(E-\omega_k) \frac{1}{2 E^{*2} \omega_k}-\frac{1}{2 \omega_k^2}
\\ & \hspace{50pt} + \frac{(E-\omega_k)}{4 E^* \omega_k^2} \left [ \frac{ [(E - \omega_k)^2 - \omega^2_{P-k} ] 2E^* }{(E - \omega_k + \omega_{P-k} )(p^{*2}-\boldsymbol k^{*2})} - \frac{\omega_k (E^* - 2 \omega_k^*) }{(p^{*2}-\boldsymbol k^{*2})} \right ] \,,
\end{split}
\\[5pt]
\begin{split}
& \to
(E-\omega_k) \frac{1}{2 E^{*2} \omega_k}-\frac{1}{2 \omega_k^2}
\\ & \hspace{70pt} + \frac{(E-\omega_k)}{4 E^* \omega_k^2} \left [ \frac{ 8E^{*2} }{(E - \omega_k + \omega_{P-k} ) (E^* + 2 \omega_k^*)} - \frac{ 4 \omega_k }{ E^* + 2 \omega_k^* } \right ] \,,
\end{split}
\end{align}
where in the second line we have canceled the poles. At this stage we can simplify to the desired result:
\begin{align}
\mathcal Q_{\boldsymbol k}(\boldsymbol P, \boldsymbol k) & \to
(E-\omega_k) \frac{1}{2 E^{*2} \omega_k}-\frac{1}{2 \omega_k^2} + \frac{(E-\omega_k)}{4 E^* \omega_k^2} \left [ \frac{ 2E^{*} }{(E - \omega_k ) } - \frac{ 2 \omega_k }{ E^* } \right ] + O(a^2) \,,
\\ & \to O(a^2) \,,
\end{align}
as claimed.

We thus find that $\boldsymbol F_a$ reduces to
\begin{align}
L_a \boldsymbol{F}_a R_a &=-\frac{1}{4 E^*}\left[\frac{1}{L^3}\sum_{\vec k} \frac{\omega_k^*}{\omega_k}-\int \frac{d^3 \vec{k}^*}{(2\pi)^3} \right] \frac{{\sf L}_a(\boldsymbol P, \boldsymbol k) \ {\sf H}(\boldsymbol k^{*}) \ {\sf R}_a(\boldsymbol P, \boldsymbol k)}{p^{*2}-\boldsymbol k^{*2}-a^2 \f_P(\boldsymbol k^*) + i \epsilon} \,,
\\[5pt]
\f_P(\boldsymbol k^*) & = \frac12 \left [ \frac{E - \omega_k}{E - \omega_{P-k}} \f(\boldsymbol k) + \f(\boldsymbol P-\boldsymbol k) \right ] \,, \nonumber
\end{align}
where we have repeated the definition of $\f_P(\boldsymbol k^*)$ for convenience.

The final steps closely resemble the analysis already performed for the zero-momentum case. First we rewrite the functions ${\sf L}_a(\boldsymbol P, \boldsymbol k)$ and ${\sf R}_a(\boldsymbol P, \boldsymbol k)$ in terms of the $\boldsymbol k^*$ as
\begin{align}
{\sf L}^{\!*}_a(\boldsymbol P,\vec{k}^{*2},\hat{\vec k}^*) = {\sf L}_a(\boldsymbol P, \boldsymbol k) \,, \qquad \qquad
{\sf R}_a^*(\boldsymbol P,\vec{k}^{*2},\hat{\vec k}^*) = {\sf R}_a(\boldsymbol P, \boldsymbol k) \,,
\end{align}
and then decompose in spherical harmonics
\begin{align}
\begin{split}
{\sf L}^{\!*}_a(\boldsymbol P,\vec{k}^{*2},\hat{\vec k}^*)&=[{\sf L}^{\!*}_a]_{lm}(\boldsymbol P, \vec{k}^{*2})\sqrt{4\pi}|\vec{k}^*|^lY_{lm}(\hat{\vec k}^*) \,, \\[5pt]
{\sf R}^*_a(\boldsymbol P,\vec{k}^{*2},\hat{\vec k}^*)&=[{\sf R}^{*}_a]_{l'm'}(\boldsymbol P, \vec{k}^{*2})\sqrt{4\pi}|\vec{k}^*|^{l'}Y^*_{l'm'}(\hat{\vec k}^*) \,,
\end{split}
\end{align}
as in eq.~\eqref{eq:first_decomp}.

These functions can now be evaluated at the pole position via
\begin{align}
\vec{k}^{*2}\to p^{*2} -a^2 \f_P(\boldsymbol k^*) \,.
\end{align}
The replacement introduces additional dependence on both $\vec{k}^{*2}$ and $\hat{ \boldsymbol{k}}^*$, and the latter is decomposed via a second set of spherical harmonics as in eq.~\eqref{eq:second_decom_LR}:
\begin{align}
[{\sf L}^{\!*}_a]_{lm} \big (\boldsymbol P, p^{*2} -a^2 \f_P(\boldsymbol k^*) \big ) & = [{\sf L}^{\!*}_a]_{lm,\alpha\beta}(\boldsymbol P, p^{*2},\vec{k}^{*2})\sqrt{4\pi}|\vec{k}^*|^\alpha Y_{\alpha,\beta}(\hat{\vec{k}}^*) \,,\\[5pt]
[{\sf R}^*_a]_{l'm'} \big (\boldsymbol P, p^{*2} -a^2 \f_P(\boldsymbol k^*) \big )&=[{\sf R}^*_a]_{l'm',\alpha'\beta'}(\boldsymbol P, p^{*2},\vec{k}^{*2})\sqrt{4\pi}|\vec{k}^*|^{\alpha'} Y^*_{\alpha'\beta'}(\hat{\vec{k}}^*) \,.
\end{align}

At this stage, the additional $\vec{k}^{*2}$ dependence appears only at $a^2$ and hence, to the order we work, we can set $\vec{k}^{*2}\to p^{*2}$. This does not introduce any additional angular dependence and hence there is no need for an additional decomposition. Using the shorthand $[{\sf L}^{\!*}_a]_{lm,\alpha\beta}(\boldsymbol P,p^{*2})=[{\sf L}^{\!*}_a]_{lm,\alpha\beta}(\boldsymbol P,p^{*2},p^{*2})$ and similar for ${\sf R}_a$, we finally reach the main result of the derivation
\begin{align}
L_a \boldsymbol{F}_a R_a=[{\sf L}^{\!*}_a]_{lm,\alpha\beta}(\boldsymbol P,p^{*2}) \, [\boldsymbol{F}_{\!a}]_{lm,l'm',\alpha\beta,\alpha'\beta'}(E, \boldsymbol P, L) \, [{\sf R}^{*}_a]_{l'm',\alpha'\beta'}(\boldsymbol P,p^{*2}) \,,
\end{align}
where
\begin{multline}
[\boldsymbol{F}_{\!a}]_{lm,l'm',\alpha\beta,\alpha'\beta'}(E, \boldsymbol P, L) =-\frac{1}{4 E^*}\left[\frac{1}{L^3}\sum_{\vec k} \frac{\omega_k^*}{\omega_k}-\int \frac{d^3 \vec{k}^*}{(2\pi)^3} \right]
\\
\times
\frac{16 \pi^2 \vert \boldsymbol k^* \vert^{l+l'+\alpha+\alpha'} Y_{lm}(\hat {\boldsymbol k}^*) Y_{\alpha \beta}(\hat {\boldsymbol k}^*) Y^*_{l'm'}(\hat {\boldsymbol k}^*) Y^*_{\alpha' \beta'}(\hat {\boldsymbol k}^*) \, {\sf H}(\boldsymbol k^{*}) }{p^{*2}-\boldsymbol k^{*2}-a^2 \f_P(\boldsymbol k^*) + i \epsilon} \,.
\end{multline}
This completes the construction of the finite-volume function for non-zero momentum in the finite-volume frame.

Using this in eq.~\eqref{eq:final_QC} gives the moving-frame version of the discretized quantization condition:
\begin{align}
\det_{lm, \alpha\beta} \left [\mathcal{M}_{a,P}(E^*)^{-1}-\boldsymbol{F}_{\!a}(E, \boldsymbol P, L) \right ] = 0 \,,
\end{align}
which depends on a version of the discretized scattering amplitude $[\mathcal{M}_{a,P}]_{l'm',lm,\alpha'\beta',\alpha\beta}(E^*)$, a moving frame generalization of the quantity introduced in eq.~\eqref{eq:final_M_rest}.

As with the zero-momentum case, the explicit construction of $[\mathcal{M}_{a,P}]_{l'm',lm,\alpha'\beta',\alpha\beta}(E^*)$ follows via a standard time-ordered correlator, defined in the Symanzik Effective Theory:
\begin{equation}
C_{a,P}(k_{\sf in},k_{\sf out})=\int \! d^4x \int \! d^4y \int \! d^4x' \, e^{i k_{\sf in} \cdot x + i (P-k_{\sf in}) \cdot y-i k_{\sf out} \cdot x'}\langle 0|{\text T} \! \left[\phi(x)\phi(y)\phi(x')\phi(0)\right]|0\rangle_a \,.
\end{equation}
One then amputates with the version of the propagator that includes $O(a^2)$ discretization effects, defined in eq.~\eqref{eq:prop_final_latt}:
\begin{align}
D_a(k)=\frac{\Z(\vec{k})}{k_4^2+ \boldsymbol k^2 + m_\pi^2 + a^2 {\sf f}(\boldsymbol k)} \,,
\end{align}
where ${\sf Z}_a(\boldsymbol k)$ is defined in eq.~\eqref{eq:residue}. The correctly normalized scattering amplitude is then defined as
\begin{equation}
\mathcal{M}_{a,P}(k_{\sf in}, k_{\sf out}) = \sqrt{{\sf Z}_a(\vec{k}_{\sf in}) {\sf Z}_a(\vec{P} - \vec{k}_{\sf in}) {\sf Z}_a(\vec{k}_{\sf out}) {\sf Z}_a(\vec{P} - \vec{k}_{\sf out})} \ C^{\sf amp}_{a,P}(k_{\sf in}, k_{\sf out}) \,.
\end{equation}

To conclude the construction we set $k_{4,{\sf in}} = i \Omega_{k_{\sf in}}$ and $k_{4,{\sf out}} = i \Omega_{k_{\sf out}}$. The remaining dependence is on the total four-momentum $P$ as well as $\vec{k}_{\sf in}$ and $\vec{k}_{\sf out}$. Following the steps above, these are then expressed in the CMF frame to write
\begin{multline}
\mathcal{M}_{a,P}(\vec{k}_{\sf in}^{*2},\hat{\vec k}_{\sf in}^*,\vec{k}_{\sf out}^{*2},\hat{\vec k}_{\sf out}^*)=
\\
[\mathcal{M}_{a,P}]_{l'm',lm}(\vec{k}_{\sf in}^{*2},\vec{k}_{\sf out}^{*2})\ 4\pi |\vec{k}^*_{\sf in}|^l|\vec{k}^*_{\sf out}|^{l'} Y^*_{l'm'}(\hat{\vec{k}}_{\sf in}^*)Y_{lm}(\hat{\vec{k}}_{\sf out}^*) \,.
\end{multline}
Finally, the on-shell replacement $\boldsymbol k^{*2} \to p^{*2}-a^2 \f_P(\boldsymbol k^*)$ is applied to the partial-wave components, followed by a subsequent decomposition in spherical harmonics
\begin{align}
[\mathcal{M}_{a,P}]_{l'm',lm}(\vec{k}^{*2}_{\sf in},\vec{k}^{*2}_{\sf out})& \to [\mathcal{M}_{a,P}]_{l'm',lm}(p^{*2}-a^2 \f(\boldsymbol k^*_{\sf in}),p^{*2}-a^2 \f(\boldsymbol k^*_{\sf out}))\\[5pt]
& \hspace{-60pt} = [\mathcal{M}_{a,P}]_{l'm',lm',\alpha'\beta',\alpha\beta}(p^{*2},\vec{k}^{*2}_{\sf in},\vec{k}^{*2}_{\sf out})4\pi |\vec{k}^*_{\sf in}|^\alpha|\vec{k}^*_{\sf out}|^{\alpha'} Y^*_{\alpha'\beta'}(\hat{\vec{k}}_{\sf in}^*)Y_{\alpha\beta}(\hat{\vec{k}}_{\sf out}^*) \,.
\nonumber
\end{align}
Setting $\vec{k}^{*2}_{\sf in} = \vec{k}^{*2}_{\sf out}= p^{*2}$ completes the construction.

\bibliographystyle{JHEP}
\bibliography{ref.bib}

\end{document}